\documentclass[prx,twocolumn,tightenlines,aps,showpacs,superscriptaddress,longbibliography,nofootinbib]{revtex4-1}

\pdfoutput=1
\usepackage[pdftex]{graphicx}

\usepackage{dcolumn}
\usepackage{bm}
\usepackage{epsfig}
\usepackage{latexsym} 
\usepackage{amsmath}
\usepackage{amssymb}
\usepackage{color}
\usepackage{array}
\usepackage{bbm}
\usepackage{color}
\usepackage{array}
\usepackage{cancel}
\usepackage[dvipsnames]{xcolor}

\usepackage[colorlinks=true,allcolors=blue]{hyperref}%

\usepackage{titlesec}
\titleformat{\paragraph}[runin]
        {\bfseries}
        {}
        {0.0em}
        {}
        [ -- ~]
\titlespacing*{\paragraph}{0pt}{4pt}{0pt}

\usepackage[normalem]{ulem} 

\usepackage{booktabs}
\newcolumntype{C}[1]{>{\centering\arraybackslash}m{#1}}
\AtBeginDocument{
\heavyrulewidth=.08em
\lightrulewidth=.05em
\cmidrulewidth=.03em
\belowrulesep=.65ex
\belowbottomsep=0pt
\aboverulesep=.4ex
\abovetopsep=0pt
\cmidrulesep=\doublerulesep
\cmidrulekern=.5em
\defaultaddspace=.5em
}
\usepackage{booktabs}
\newcolumntype{R}[1]{>{\raggedleft\arraybackslash}p{#1}}
\AtBeginDocument{
\heavyrulewidth=.08em
\lightrulewidth=.05em
\cmidrulewidth=.03em
\belowrulesep=.65ex
\belowbottomsep=0pt
\aboverulesep=.4ex
\abovetopsep=0pt
\cmidrulesep=\doublerulesep
\cmidrulekern=.5em
\defaultaddspace=.5em
}

\newcommand{\<}{\langle}
\newcommand{\e}{\varepsilon}

\renewcommand{\>}{\rangle}
\renewcommand{\(}{\left(}
\renewcommand{\)}{\right)}
\renewcommand{\[}{\left[}
\renewcommand{\]}{\right]}
\renewcommand{\v}[1]{\boldsymbol{#1}} 

\newcommand{\eps}{\epsilon}

\newcommand{\U}{\mathcal{U}}

\newcommand{\Z}{\mathbb{Z}}

\newcommand{\red}[1]{\textcolor{red}{#1}}

\newcommand{\skippage}{\onecolumngrid \newpage \twocolumngrid}

\begin{document}
\title{Realizing a dynamical topological phase in a trapped-ion quantum simulator}
\author{Philipp T. Dumitrescu}
\email{pdumitrescu@flatironinstitute.org}
\affiliation{Center for Computational Quantum Physics, Flatiron Institute, 162 5th Avenue, New York, NY 10010, USA}

\author{Justin Bohnet}
\affiliation{Honeywell Quantum Solutions, 303 S. Technology Ct., Broomfield, Colorado 80021, USA}

\author{John Gaebler}
\affiliation{Honeywell Quantum Solutions, 303 S. Technology Ct., Broomfield, Colorado 80021, USA}

\author{Aaron Hankin}
\affiliation{Honeywell Quantum Solutions, 303 S. Technology Ct., Broomfield, Colorado 80021, USA}

\author{David Hayes}
\affiliation{Honeywell Quantum Solutions, 303 S. Technology Ct., Broomfield, Colorado 80021, USA}

\author{Ajesh Kumar}
\affiliation{Department of Physics, University of Texas at Austin, Austin, TX 78712, USA}

\author{Brian Neyenhuis}
\affiliation{Honeywell Quantum Solutions, 303 S. Technology Ct., Broomfield, Colorado 80021, USA}

\author{Romain Vasseur}
\affiliation{Department of Physics, University of Massachusetts, Amherst, MA 01003, USA}

\author{Andrew C. Potter}
\email{acpotter@utexas.edu}
\affiliation{Department of Physics, University of Texas at Austin, Austin, TX 78712, USA}
\affiliation{Department of Physics and Astronomy, and Quantum Matter Institute,
University of British Columbia, Vancouver, BC, Canada V6T 1Z1}

\begin{abstract}
Nascent platforms for programmable quantum simulation offer unprecedented access
to new regimes of far-from-equilibrium quantum many-body dynamics in (approximately) isolated systems. Here, achieving precise control over quantum many-body entanglement is an essential task for quantum sensing and computation. Extensive theoretical work suggests that these capabilities can enable dynamical phases and critical phenomena that exhibit topologically-robust methods to create, protect, and manipulate quantum entanglement that self-correct against large classes of errors. However, to date, experimental realizations have been confined to classical (non-entangled) symmetry-breaking orders~\cite{zhang2017observation,choi2017observation,kyprianidis2021observation}. In this work, we demonstrate an emergent dynamical symmetry protected topological phase (EDSPT)~\cite{friedman2020topological}, in a quasiperiodically-driven array of ten $^{171}\text{Yb}^+$ hyperfine qubits in Honeywell's System Model H1 trapped-ion quantum processor~\cite{pino2021demonstration}. This phase exhibits edge qubits that are dynamically protected from control errors, cross-talk, and stray fields. Crucially, this edge protection relies purely on emergent dynamical symmetries that are absolutely stable to generic coherent perturbations. This property is special to quasiperiodically driven systems: as we demonstrate, the analogous edge states of a periodically driven qubit-array are vulnerable to symmetry-breaking errors and quickly decohere. Our work paves the way for implementation of more complex dynamical topological orders~\cite{harper2020topology} that would enable error-resilient techniques to manipulate quantum information.
\end{abstract}
\maketitle

Understanding and categorizing new types of universal dynamical phenomena --- the dynamical analogs of (meta)stable phases and critical phenomena --- that can arise in isolated quantum many-body systems poses a fundamental scientific challenge. Early investigations have already yielded deep insights into the quantum mechanical underpinnings of thermalization and chaos~\cite{abanin2019colloquium}, and shown how thermalization can be prevented by artificial randomness and disorder through many-body localization (MBL). MBL can protect long lived quantum coherent dynamics in ``hot", dense, and strongly-driven matter, and can enable new classes of inherently dynamical quantum phases with properties that would be fundamentally forbidden in static thermal equilibrium, such as dynamical symmetry breaking and topology~\cite{harper2020topology}. 

From a practical point of view, universal and quantum coherent dynamical behaviors tantalizingly offer error-resilient methods to create, protect, and manipulate quantum many-body entanglement --- the driving-force of quantum computation. To perform a quantum computation, one faces a trade-off between the desire to isolate qubits to preserve their coherence, and the need to strongly interact qubits in order to perform computations. Even in perfect isolation from environmental decoherence, strong inter-qubit coupling inevitably leads to residual, coherent errors --- from stray fields, gate miscalibrations, cross-talk, etc. --- that disrupt computations. Perhaps counterintuitively, coherent errors can be more damaging than incoherent ones. In particular, the infidelity resulting from $N$ gates with error-amplitude $\eps$ can grow as $\sim N^2\eps^2$ for coherent errors compared to $\sim N\eps^2$ for incoherent ones~\cite{sanders2015bounding}. Despite their outsized detrimental impact on algorithm performance, coherent errors are challenging to detect. Standard randomized benchmarking procedures, for example, combine both coherent and incoherent errors into a single effective error-per-gate, which can dramatically overestimate the accuracy of structured circuits relevant for computations.

Employing dynamical decoupling pulse sequences is a time-honored approach to mitigate certain types of coherent errors associated with uncontrolled static stray fields. However, for traditional methods using global, single-spin control, slight imperfections $\eps$ in dynamical decoupling pulses accumulate and spoil the decoupling in time $\sim 1/\eps$.  By contrast, recent progress in understanding dynamical phases~\cite{harper2020topology} has theoretically predicted that local dynamical control of \emph{multi-spin} interactions can enable self-correcting dynamical decoupling sequences that are inherently robust against large classes of coherent errors. The robustness of these schemes arises from sharply quantized topological invariants of the dynamics that cannot be altered by generic coherent perturbations below a critical strength, in direct analogy to the stability enjoyed by equilibrium phases of matter. 
 In these driving protocols, termed \emph{dynamical topological phases}, engineered random couplings (``disorder") provide a key stabilizing element yielding long-lived many-body localization that avoids thermalization and its associated chaotic scrambling of quantum information. 
 Despite extensive theoretical progress in formally classifying and theoretically characterizing dynamical topological phases in both periodically-driven (Floquet) and quasi-periodically driven systems, to date only non-topological time-crystalline phases with \emph{classical} (relying neither on entanglement or coherence) rather than \emph{quantum} dynamical orders have been achieved experimentally~\cite{zhang2017observation,choi2017observation,kyprianidis2021observation}.
 
\begin{figure*}[t!] 
    \centering
    \includegraphics[width=\textwidth]{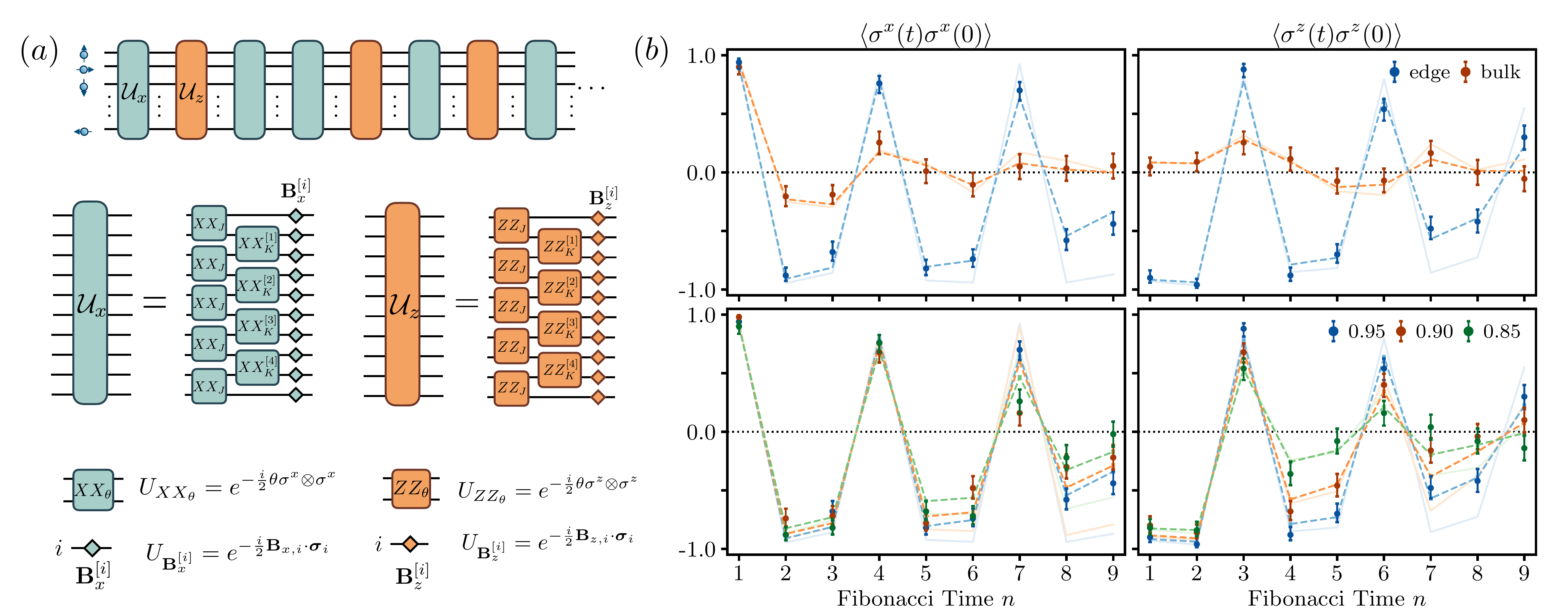}
    \caption{ {\bf Fibonacci drive (EDSPT) model}
    \label{fig:qp} 
    (a) Schematic of a Fibonacci sequence of two elementary circuit layers $\U_{x,z}$ that defines the quasiperiodic model, whose ideal implementation realizes am (exponentially long-lived, though ultimately metastable) EDSPT phase with edge states protected purely by emergent dynamical symmetries that are robust against generic quasiperiodic perturbations. Couplings $K^{[i]}$ are drawn randomly and independently for each odd-bond from $[0,4\pi]$, and on-site fields $\v{B}^{[i]}$ are drawn independently for each site, $i$ with magnitude $\sim [0,4\pi]$ and uniformly random direction, breaking all microscopic symmetries. (b) Edge and (site-averaged) bulk spin correlators sampled at Fibonacci times, $t_n = F_n$ up to maximum time $F_9=54$, for exchange coupling $J = 0.95\pi$ (top row). Pale lines show ideal (noiseless) simulations, dashed lines show simulations with depolarizing noise channel, and dots with $1\sigma$ error bars indicate experimental data. Bulk correlations rapidly decay due to random fields. Edge correlators enjoy topologically-enchanced coherence and exhibit characteristic persistent period-three oscillations in the Fibonacci index $n$. This behavior persists over a range of couplings (bottom row).
        }
\end{figure*}

\begin{figure*}[t]
    \centering
    \includegraphics[width=1.0\textwidth]{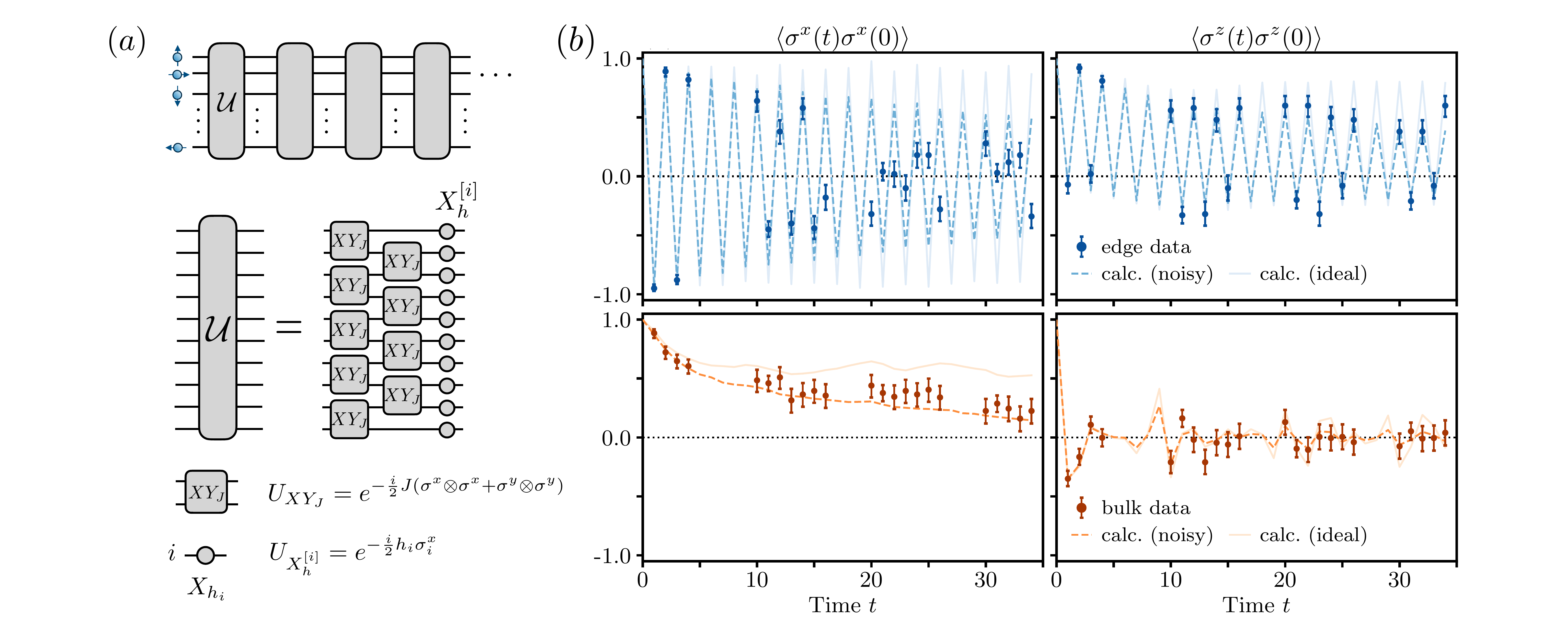}
    \caption{
    \label{fig:fspt} 
    {\bf Floquet model} 
    (a) Circuit implementation of the FSPT model proposed in~\cite{kumar2018string} generated by sequence of repeated circuit layers $\U$, each consisting of a brickwork of nearest neighbor XY-interactions followed by random fields along $x$ that are independently and identically distributed on each site $\sim [0,4\pi]$ (maximal disorder strength). For $|J-\pi|\lesssim 0.2\pi$, numerical simulations predict an FSPT phase with bulk MBL dynamics and topological edge modes protected by a combination of symmetry, topology, and dynamics. (b) Experimental spin auto correlation data (data points) atop theoretical calculations with (dashed lines) and without (solid lines) incoherent depolarizing noise. The site-averaged bulk data 
 show behavior consistent with weakly-open MBL dynamics. Though $\sigma^z$ correlations exhibit quantitative agreement with incoherent depolarizing noise simulations, the characteristic period-doubled oscillations of the edge-spin symmetric ($\sigma^x$) correlators quickly dephase after $t\gtrsim 15$ oscillations. We attribute this behavior to the presence of coherent symmetry breaking phase errors as discussed in the text. The vulnerability of the FSPT edge states to such errors highlights the comparative robustness of the EDSPT phase.
        }
\end{figure*}

In this work, we experimentally implement two models of inherently-quantum
dynamical topological phases in (quasi)periodically-driven $1d$ array of ten $^{171}\text{Yb}^+$ hyperfine spins in Honeywell's System Model H1 quantum charge-coupled device (QCCD) trapped-ion architecture~\cite{pino2021demonstration}. These two drive protocols, respectively implement i) a quasiperiodically-driven Emergent Dynamical Symmetry Protected Topological Phase~(EDSPT) illustrated and defined in Fig.~\ref{fig:qp}, and ii) a Floquet symmetry protected topological phase (FSPT) (Fig.~\ref{fig:fspt}). The Floquet model was previously introduced in~\cite{kumar2018string}, and the EDSPT model is a ``stroboscopic" version of the model studied in~\cite{friedman2020topological} that is more amenable to implementation on a gate based quantum processor (though this modification produces metastability on exponentially long-time scales beyond the experimental life-time as we discuss extensively in Appendix~\ref{AppendixMagnus}).

The hall-mark of these topological phases are robust edge-modes that can phase-coherently retain information despite strong and repeated interactions with other qubits and external fields, and which slowly decay at a rate set only by the incoherent errors and imperfect environmental isolation of the trapped-ion qubits. The dynamical topological protection of the EDSPT edge states does not require any fine-tuned symmetry, but rather stems from purely emergent dynamical symmetries that are ``absolutely stable"~\cite{von2016absolute} to generic coherent perturbations.
The quasiperiodicity of the drive is essential for achieving this property: topological order without symmetry protection is fundamentally impossible in $1d$ bosonic (spin) systems with static or periodically time-dependent (Floquet) Hamiltonians~\cite{friedman2020topological}.
In stark contrast, in the FSPT model, the edge spins rely on a fine-tuned microscopic symmetry that renders them fragile to coherent symmetry-breaking errors, which naturally arise in all quantum simulation platforms. We show that this vulnerability causes the FSPT edge spins to quickly dephase after only a handful of drive cycles. We observe that the EDSPT edge states are insensitive to the same coherent errors that destroy the FSPT edge states (and in fact even to much stronger intentionally-introduced coherent errors). 

While the FSPT realization only survives for short times, our results suggest that its periodically repeating drive protocol and symmetry sensitivity amplify coherent errors. By contrasting decay rates of bulk and boundary spin correlations along different axes (which have different sensitivities to different types of coherent and incoherent errors) these drive protocols can actually serve as a useful tool for diagnosing coherent errors along selectable channels. 
Moreover, in Appendix~\ref{app:fluxecho} we demonstrate a novel many-body interferometric probe to detect the dynamical topological order of the FSPT (which cannot be detected by any local measurements).

\paragraph{Weakly-open MBL}
Before discussing results, we briefly remark on the expected behavior. MBL systems are defined by the emergence of an extensive number of local conservation laws, and associated local integrals of motion (LIOMs)~\cite{abanin2019colloquium}.
In an idealized, perfectly-isolated MBL system, site- or disorder- averaged spin correlations would exhibit a rapid, transient decay before saturating to a long-time value set by the overlap of the single-site spin operator with the LIOMs (which may vanish if dictated by symmetry). In practice, however, any experiment is partially open to its environment resulting in a slow melting of MBL and gradual decay of non-thermal correlations.  
We will refer to MBL-like dynamics that decay on a time scale set by gate-error rates (generally much longer than the interaction time-scale), as \emph{weakly-open MBL}. We will contrast the observed behavior to noisy simulations with depolarizing noise channels with one- and two-qubit gate depolarization parameters $p_{1q},p_{2q}$ respectively, which we estimate from randomized benchmarking experiments. We note that, the weakly-open MBL dynamics is \emph{not} sensitive to weak coherent errors, since MBL is stable to generic unitary-preserving perturbations, these only weakly perturb the LIOM structure and give minor quantitative corrections to correlations.

\paragraph{Topology from Quasiperiodic Driving}  The EDSPT model consists of a Fibonacci sequence of two different types of circuit layers $\U_x, \U_z$~\cite{dumitrescu2018logarithmically} as defined and illustrated in Fig.~\ref{fig:qp}(a). Specifically, the sequence of unitaries at ``Fibonacci times" (i.e.~for number of circuit layers, $t_n=F_n$ where $F_n$ is the $n^\text{th}$ Fibonacci number defined by $F_{n+1}=F_{n-1}+F_{n}$, $F_{0,1}=1$) is defined recursively through, $\U(t=F_n) = \U_n$, with:
\begin{align}
	\U_{n+1} = \U_{n-1}\U_n, ~~~~ \U_1=\U_x,~~\U_2 = \U_z\U_x
\end{align}
This recursion relation generates a quasiperiodic sequence of time-dependent unitaries. We emphasize that the random axis fields $\v{B}^\alpha$ in each layer completely break all microscopic symmetries~\footnote{In Appendix~\ref{app:additionaldata}, we also implement an analogous drive with a fine-tuned Ising symmetry axis analogous to the FSPT model studied below, and find similar results.}.

This model represents a discrete-pulse (``stroboscopic") version of the smooth drive that was examined in \cite{friedman2020topological} through numerical simulations and analytic techniques of~\cite{else2020long}. There, it was shown that for pulse strength $J$ close to $\pi$, the system exhibited a pair of emergent dynamical $\Z_2$ symmetries despite lacking any true microscopic symmetries. This model exhibits edge states characterized by the same topological invariant as the equilibrium AKLT/Haldane spin chain~\cite{affleck2004rigorous,haldane1983nonlinear,chen2012symmetry,pollmann2012symmetry}, but with the crucial difference that this behavior is not fine-tuned to symmetric drives, but rather automatically self-corrects against all generic coherent perturbations to the drive. 

The characteristic behavior of the topological regime of this model are topological edge states that undergo coherent quasiperiodic oscillations, which can be most easily understood by examining their behavior at Fibonacci times, $t_n=F_n$, where the edge motion exhibits $3n$-periodic oscillations~\cite{dumitrescu2018logarithmically,friedman2020topological} indicative of the periodicity of the even/odd parity of Fibonacci numbers. In the experimental implementation, we observe weakly-open MBL versions of these topological edge-oscillations (see Fig.~\ref{fig:qp}) over a wide range of pulse strengths $J$. The data matches quantitatively to simulations using depolarizing noise channels with randomized benchmarking (RBN)-measured error rates, despite the presence of coherent gate errors and intentional breaking of all symmetries. These results highlight the robustness of the emergent dynamical symmetries protecting the EDSPT  edge-modes against generic coherent errors.

While the pulsed version implemented here is much more convenient for implementation on a gate-based quantum simulator, avoiding the need to discretize a smooth time-dependence at significant cost to circuit depth and error accumulation, it does introduce some (in principle) important changes in the resulting dynamics~\cite{dumitrescu2018logarithmically,else2020long}. Namely: even in a perfectly-isolated system, the recursive drives would exhibit logarithmically slow heating~\cite{dumitrescu2018logarithmically} with numerical simulations showing that MBL dynamics eventually melting away into infinite temperature incoherent at an exponentially-long `heating' time-scale $t_h\sim e^{1/\delta}$~\cite{dumitrescu2018logarithmically} where $\delta$ quantifies the deviation from an ideal (purely commuting) drive (see Appendix~\ref{AppendixMagnus} for detailed definitions). In Appendix \ref{AppendixMagnus}, we adapt the recursive Magnus expansion of~\cite{dumitrescu2018logarithmically} to treat topologically non-trivial drives, and demonstrate analytically the presence of emergent dynamical symmetries lasting up to parametrically long time scales $t_*\sim \delta^{-3}$, and provide numerical evidence that the emergent symmetries and topological edge dynamics actually survive all the way to $t_h\gg t_*$. We emphasize that this exponential dependence makes this metastability irrelevant in practice -- as $t_h$ can readily be pushed far beyond the experimental lifetime by modest changes in $\delta$.

\paragraph{An ersatz FSPT} 
The EDSPT model above provides an example of quasiperiodically driven spin chain which exhibits robust edge states that do not rely on symmetry protection. To emphasize the importance of this feature, we now turn to a Floquet model which shows that symmetry-protection requirements render topological edge states vulnerable to coherent errors (see Fig.~\ref{fig:fspt}).
An ideal implementation of this drive respects a $\Z_2$ (Ising) symmetry generated by $\pi$ rotations about $x$: $g=\prod_i \sigma^x_i$. This Floquet model was previously introduced by some of us~\cite{kumar2018string} in the context of studying non-local string order parameters for the FSPT phase. When the exchange coupling $J$ is close to $\pi$: $|J/\pi-1|<\delta_c\approx 0.2$, numerical simulations~\cite{kumar2018string} predict an MBL FSPT phase with bulk MBL dynamics and topological edge states (for open boundary conditions). When the protecting symmetry is intact, the edge modes undergo coherent period-two oscillations that dynamically decouple them from bulk degrees of freedom despite repeated strong interactions between all neighboring spins. Formally, the topological dynamics of the edge is characterized by local anticommuting action of time-translation and Ising symmetry generators, corresponding to a non-trivial group cohomology element of $\mathcal{H}^2\(\mathbb{Z}\times\mathbb{Z}_2,U(1)\)=\Z_2$~\cite{von2016phase,else2016classification,potter2016classification,roy2016abelian}, which can be physically understood as a quantized pumping of Ising symmetry ``charge" (i.e.~parity quantum number for $\pi$-rotations about $x$) onto the boundary in each driving period~\cite{von2016phase,else2016classification,potter2016classification,roy2016abelian,kumar2018string}. In Appendix~\ref{app:fluxecho}, we exploit this pumping picture to devise and implement interferometric probe to directly measure this topological invariant.

Figure~\ref{fig:fspt} shows experimental spin-correlations data for this model with $J=0.9\pi$ up to $35$ Floquet periods (see also Appendix~\ref{app:additionaldata} for additional parameters), along side noisy and ideal simulations. The (site-averaged) bulk correlators exhibit characteristic short-time transient decay followed by slowly-decaying plateau for symmetric correlators ($C_x$) and near-zero value for ($C_z$), consistent with weakly-open  and symmetry-preserving MBL dynamics, and quantitatively matches simulations with depolarizing noise with RBN-determined parameters.

By contrast, the edge spins exhibit weakly-open period-doubled amplitude oscillations ($C_z$) but which quickly dephase as indicated by the random behavior of $C_x$ for $t\gtrsim 15$ Floquet periods. This behavior cannot be explained by incoherent errors alone, but rather is consistent with significant $2-3\%$ coherent error amplitudes per two-qubit gate that appear to be predominately phase errors that affect the $C_x$ but not $C_z$ edge correlators. We note that these errors translate to much smaller gate infidelities of $\sim 5\times 10^{-4}$ consistent with randomized benchmarking chracterizations~\cite{pino2021demonstration}.
We discuss possible physical origins for this error in Appendix~\ref{AppendixErrors}, and suggest that these coherent errors most likely stem from small inhomogeneities and drifts in magnetic field that can accumulate up to $\approx 10^\circ$ rotations about $\sigma^z$ per Floquet period on each qubit, and which are stable over the time-scale of individual circuits. These account qualitatively and quantitatively for the observed dephasing of the FSPT edge states. While this error mechanism is specific to trapped-ion qubits, similar types of coherent errors are pervasive across other hardware platforms.

These results highlight that the symmetry protection requirement for FSPT edge states is a significant liability in practice, as discrete protecting symmetries are fine tuned and broken by inevitable coherent control and calibration errors. Interestingly, while these coherent errors are difficult to detect by other means, the different sensitivity of bulk and edge correlators and symmetry-axis resolved edge behavior give characteristic fingerprints of the magnitude and symmetry-structure of coherent errors. The question of whether other MBL (F)SPTs with more complex symmetry groups could be used to diagnose general multi-qubit coherent error channels is a potentially interesting line for future inquiry.

\paragraph{Discussion} 
The EDSPT  implemented here represents the first experimental realization of a purely dynamical topological phase that cannot arise in equilibrium, and of a one-dimensional bosonic topological phase that does not rely on symmetry protection. As we have demonstrated, the latter quality makes the resulting edge-state phenomenology considerably more robust than its more fragile SPT and FSPT counterparts, and opens the door to new strategies for dynamically extending coherent information storage in the presence of strong interactions and cross-talk between many-qubits. Higher-dimensional dynamical topological models have been predicted to offer even more dramatic capabilities to manipulate entanglement in a manner that is robust against coherent errors, enabling for example chiral transfer of quantized packets of quantum information~\cite{po2016chiral}, and error-resilient implementation of non-transversal operations on logical qubits encoded in the boundary of topological codes~\cite{po2017radical}. Our work paves the way for harnessing these capabilities for practical quantum information processing. 

\vspace{4pt}\noindent {\it Acknowledgements --} 
We thank Yuxuan Zhang and Michael Foss-Feig for helpful discussions, as well as
Dominic Else, Aaron Friedman, Wen Wei Ho, and Brayden Ware for prior
collaboration on this topic. We thank the entire Honeywell Quantum Solutions
team for their many contributions. This work was supported by NSF Convergence
Accelerator Track C award 2040549 (ACP),  the US Department of Energy, Office
of Science, Basic Energy Sciences, under Early Career Award No.~DE-SC0019168
(RV), and the Alfred P. Sloan Foundation through Sloan Research Fellowships (RV
and ACP). The Flatiron Institute is a division of the Simons Foundation.
Numerical simulations were performed in part on the Lonestar5 supercomputing
system at the Texas Advanced Computing Center (TACC) at UT Austin.

\section*{Methods}
\paragraph{Honeywell's System Model H1 QCCD Architecture}
Experiments were performed on Honeywell's System Model H1 trapped-ion quantum processor~\cite{pino2021demonstration} based on a Honeywell-fabricated planar chip trap operating with three parallel gate zones and 10 qubit ions. Qubits are encoded in two clock states: $\{|0\> = |F=0,m_F=0\>,|1\>=|F=1,m_F=0\>\}$ of the S$_{1/2}$ hyperfine manifold of $^{171}\text{Yb}^+$ ions, where $F,m_F$ are respectively the total internal angular momentum and projection onto a $\approx 5$G magnetic field axis. Accompanying, co-trapped $^{138}\text{Ba}^+$ ions are used for sympathetic cooling of ion motional modes without affecting logical states. Ions are trapped in either single-qubit (1q) Yb-Ba or two-qubit (2q) Yb-Ba-Ba-Yb linear ``crystal" configurations, which can be transported, orientation-swapped, split ($2q\rightarrow \{1q,1q\}$), or combined  ($ \{1q,1q\}\rightarrow 2q$) using an array of electrodes to achieve arbitrary pairings of qubit ions. During transport, the qubit logical states are essentially perfectly decoupled from their motion. Laser-based logical 1q and 2q gates are performed in parallel across three gate zones with typical infidelities of: $p_{1q}\approx 10^{-4}$ and $p_{2q}\approx 3-5\times 10^{-3}$ determined by randomized benchmarking (RBN). 1q gates implement arbitrary amplitude rotations about arbitrary axis in the $\sigma^{xy}$-plane, while $\sigma^z$ rotations are implemented virtually by updating laser phases of future 1q gates. The native entangling 2q gate is a M{\o}lmer-S{\o}rensen (MS) gate wrapped with single-qubit dressing pulses to achieve a phase-insensitive operation $u_\text{MS}=\exp\[-i\frac\pi4 \sigma^z\otimes\sigma^z\]$.

\begin{figure}[t!] 
    \centering
    \vspace{14pt}
    \includegraphics[width=0.35\textwidth]{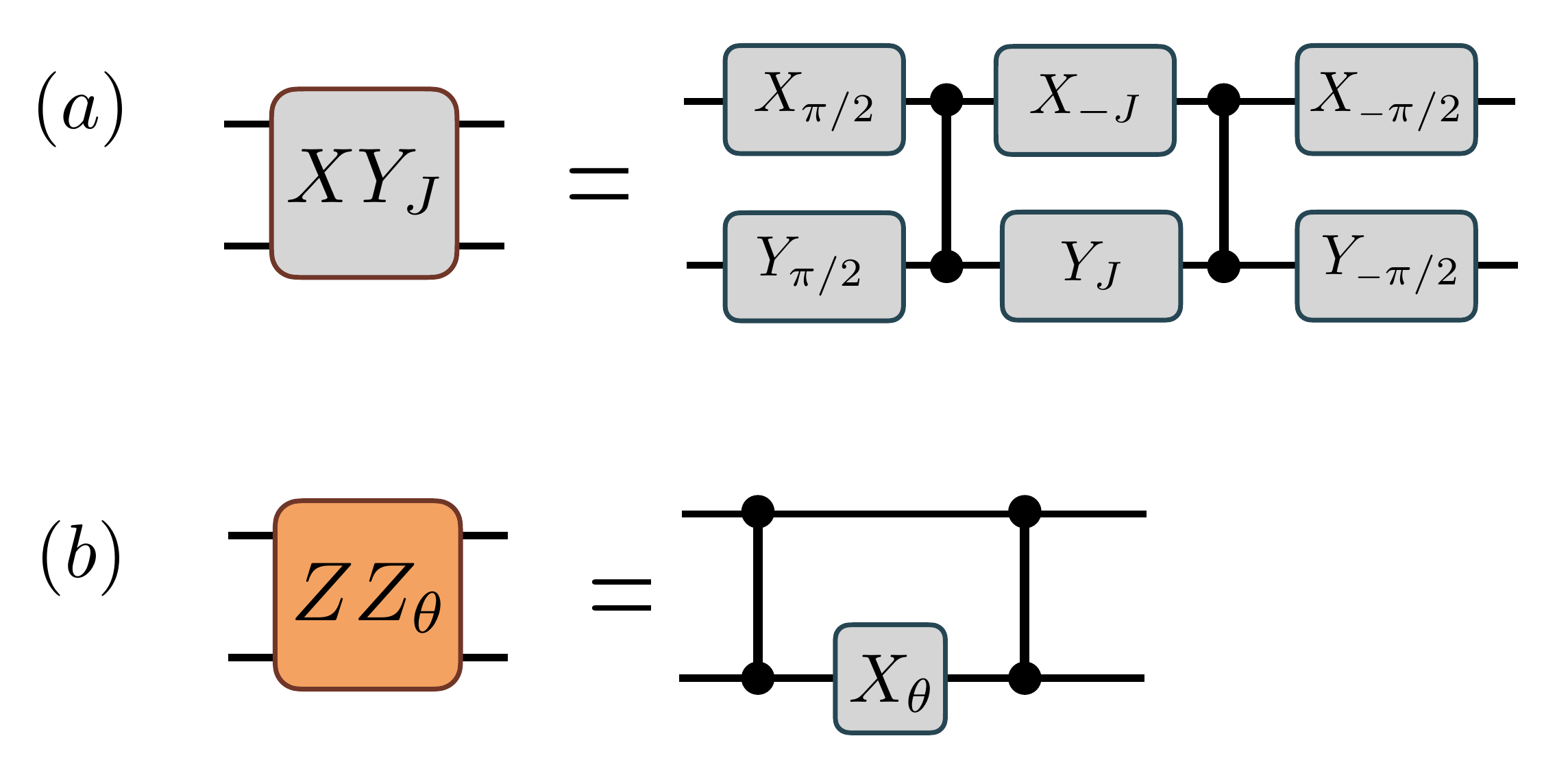}
    \caption{ \label{suppfig:circuits} 
    {\bf Circuit Compilation} of two-qubit operations into hardware-native gates for (a) the FSPT model (Fig.~\ref{fig:fspt}), and (b) the EDSPT model (Fig.~\ref{fig:qp}). $XX_\theta$ gates are obtained from a single-qubit basis rotation of $ZZ_\theta$ gates. Here, a line with two dots indicates a controlled-Z gate with unitary: $u_{C_Z} = e^{-i\pi/4(1+\sigma^z)\otimes (1+\sigma^z)}$, and single qubit gates are listed as Pauli rotations with unitary $u_{X,Y_{\theta}}=e^{-i\frac\theta2 \sigma^{x,y}}$. }
\end{figure}

\paragraph{State Preparation and Measurement} 
In each experimental implementation, we perform circuit-based time evolution up to time $t$ with unitary $U(t)$ (defined in Figs.~\ref{fig:qp},\ref{fig:fspt}) acting on $L=10$ qubits, and measure spin auto-correlations:
\begin{align}
C_\alpha(r,t) = \overline{\<\sigma^\alpha(r,t)\sigma^\alpha(r,0)\>},
\end{align}
where $\sigma^{\alpha}(r)$ ($\alpha\in\{x,y,z\}$) are standard Pauli matrices on site $r$ and $\sigma^\alpha(r,t) = U^\dagger(t)\sigma^\alpha(r,0)U^{\vphantom\dagger}(t)$ denotes the corresponding (Heisenberg picture) time evolved operator. Here, $\overline{\(\dots\)}$ denotes averaging over $N_s=100$ ``shots" with a different random initial state in each shot. Only the sample error from finite $N_s$ is included in the 1$\sigma$ error bars shown in the figures. Due to the low-clock rate of the QCCD architecture (each experiment takes between $1-2.5$ seconds per shot), we employ two tricks to obtain adequate statistical accuracy. First, we  prepare the initial states by randomly initializing product states with $\sigma^z\in\{\pm 1\}$ on even sites and $\sigma^x\in\{\pm 1\}$ on odd sites, so that measurements of $C_{x}$ and $C_{z}$ can be conducted in parallel (since there is statistically no difference between even and odd sites). Second, in each model, we focus on a single disorder realization, but verify with extensive classical simulations that our results are indicative of the generic behavior of the disorder ensemble.

\paragraph{Circuit Compilation and simulation}
The compilation of the two-qubit circuit elements for implementing the FSPT and EDSPT  into native gates are shown in Fig.~\ref{suppfig:circuits}. Each requires a pair of native $2q$ gates (shown here as control-Z gates, which differ from the MS gate only by single qubit dressing).

Noisy circuit simulations are performed with Qiskit~\cite{abraham2019qiskit} using a featureless depolarizing noise channel with depolarizing probabilities $p_{1q}=5\times 10^{-4}$, $p_{2q}=8\times 10^{-3}$. Though this simple depolarizing noise is not hardware realistic (e.g., ignores coherent errors), it provides a useful point of comparison, since departures of experimental and simulated data signal the presence of structured noise.

\bibliography{edspt_bib}

\skippage
\appendix

\section{Diagnosing coherent errors}\label{AppendixErrors}
The System Model H1 QCCD architecture has a variety of different coherent error sources arising from drifts or miscalibrations in gate-laser amplitude or phase or spatial variations in magnetic field. 
We expect that the two-qubit (2q) gate error sources are predominately incoherent, and focus our attention on possible sources of coherent errors in single qubit (1q) gate operations and idle/memory errors.

The FSPT model provides multiple characteristic finger-prints that enable us to narrow down the physical mechanism for the observed coherent errors. The circumstantial evidence is summarized as follows: first, the coherent error in question has a strong anisotropy affecting $\sigma^x$ correlators but not $\sigma^z$ correlators. Second, the amplitude of the error must be such that it can accumulate $\sim \pi$ phase within $\approx 10-15$ Floquet periods. Third, the coherent error effects are much more pronounced for the periodic-boundary condition (PBC) circuits studied in Appendix~\ref{app:fluxecho}, where they cause strong deviations from weakly-open MBL behavior in $\approx 5$ Floquet periods, compared to the $\approx 10-15$ Floquet periods for open boundary condition (OBC) circuits presented in the main text.

The last point is particularly telling since the PBC and OBC circuits have essentially the same number of one and two qubit gates, and differ mainly in the physical paths of the ions through the processor. Namely, whereas OBC circuit structures match the physical 1d layout of the H1 chip, and require much less transport of ions between different gate zones to execute, the PBC circuits require at least a pair of ions to traverse back and forth across the trap. The pronounced $\sim 2-3\times$ increase in coherent error effects strongly suggests that the dominant error source arises from different accumulation of qubit phases in different locations of the trap. 

From the variance of corrections during ($\sim$ hourly) recalibrations, we estimate that spatial variations of the laser phase or amplitude across gate zones contribute $\approx 30-50{\rm mrad}$ deviations in the axis and rotation angle of each 1q gate. Since there are $\approx 6$ 1q-gates per Floquet period, and conservatively assuming that subsequent errors add linearly, this error-source would require $\approx 10^2$ Floquet periods to accumulate a $\pi$-error: an order of magnitude too large to account for the observed data.

\begin{figure}[t!] 
    \centering
    \includegraphics[width=0.475\textwidth]{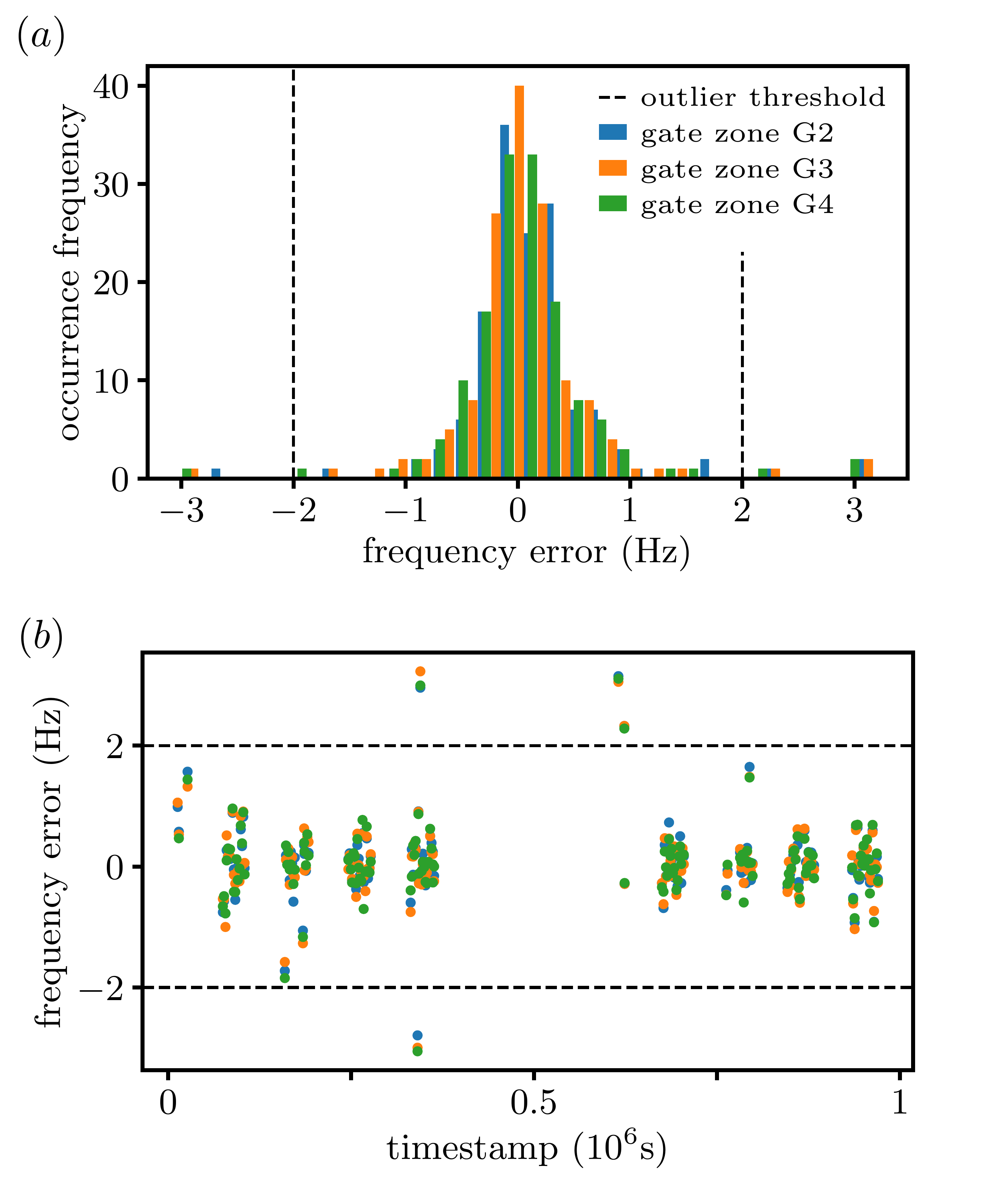}
    \caption{ {\bf Global qubit frequency calibration error}  (a) Histogram of the observed difference between the calibrated qubit frequency and the last known good calibration for each of the three ion trap gate zones: G2,G3, and G4. (b) Time-series of same data. These errors arise due to combination of background drifts in the magnetic field environment and statistical error in the actual qubit frequency calibration. For the data detailed in this paper, we observe a global RMS frequency error of $\pm {\rm 0.5(1)}$ Hz. Outliers are filtered (dashed line) for points greater than 4x the standard deviation to better capture the average behavior of this error source.
    \label{suppfig:bfield} 
        }
\end{figure}

A more serious source of coherent errors are deviations in the magnetic field, $B$ used to split off the $F=1,m_F=\pm 1$ hyperfine states from the $F=1,m_F=0$ qubit state.  From Ramsey spectroscopy, we estimate that the the $|0\>$ and $|1\>$ qubit states energies differ among gate zones with RMS variation of $\delta f_\text{RMS}\approx 0.5 {\rm Hz}$ (see Fig.~\ref{suppfig:bfield}).
These frequency offsets drift slowly over the course of many shots, but are stable over the time scale of a single circuit.
Together, these effects can be modeled by including a Hamiltonian:
\begin{align}
	H_{\Delta B} = \sum_i 2\pi ~\delta f\[x_i(t)\]\sigma^z_i
\end{align}
where $x_i(t)$ is the position of the $i^\text{th}$ ion along the linear trap axis, which has a periodic time dependence inherited from the periodicity fo the Floquet circuit.
Since each circuit layer takes on the order of $t_\text{layer}\approx 50{\rm ms}$, and given that there are two circuit layers per Floquet layer, we estimate these B-field induced memory errors can produce a $\pi$ shift in 
$N_\text{periods}\approx \pi/\(2\pi \delta f_\text{RMS} t_\text{layer}\) \gtrsim 10$ 
Floquet periods, roughly consistent with the time scale at which the FSPT implementation deviates from the simulations with purely-incoherent depolarizing noise. 
The observation that coherent error effects are exacerbated in PBC circuit compared to the OBC circuit is consistent with our observations that the frequency variations are larger between distant gate zones, since ions in the PBC circuits are transported for longer distances through the trap during each Floquet cycle.

\section{Detection of FSPT order through many-body interferometry}\label{app:fluxecho}
In this section we theoretically introduce a nonlocal ``bulk order parameter" for the FSPT phase, and exploit the flexible qubit connectivity of the QCCD architecture to measure it through a many-body interferometry. As with the local spin correlations presented in the main text, these demonstrations are limited to transient short time regimes due to the accumulation of coherent errors. 

The premise of the non-local order parameter, which we dub the ``Loschmidt flux echo" is to consider a ring with closed, periodic boundary conditions (PBC), such that there are no edge states and any topology is evident only in global features. Next, we effectively ``gauge" the protecting symmetry, i.e.~promote  the global $\Z_2$ symmetry generated by $g=\prod_i \sigma_i^x$ to a local gauge-redundancy, and explore the affect of a dynamically inserting a classical (non-fluctuating) background gauge-flux by measuring the trace-overlap of the time-evolution unitaries with- and without- a flux inserted:
\begin{align}
	\mathcal{Z}(t) = \frac{\text{tr}~\U^\dagger_F(t) \U(t)}{\text{tr} \mathbbm 1}
\end{align}
where $\U_F(t)$ denotes the unitary for time-evolution with a symmetry-flux inserted (defined in detail below), and $t$ are assumed to be integers (i.e.~multiples of the Floquet period).

\begin{figure}[t!] 
    \centering
    \includegraphics[width=0.475\textwidth]{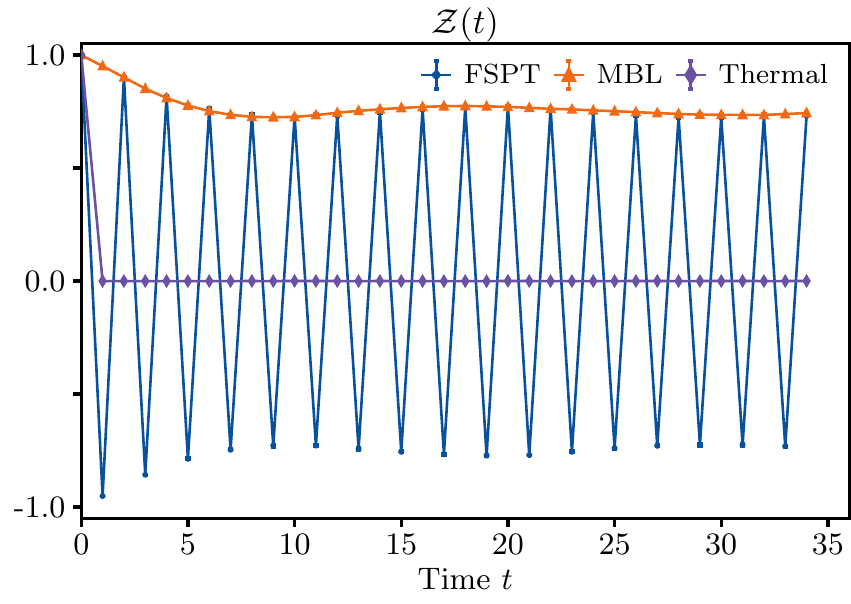}
    \caption{ {\bf Loschmidt flux echo (ED Simulations)} Exact diagonalization (ED) simulations of the Loschmidt flux echo, $\mathcal{Z}(t)$, for the Floquet model with $L=9$ for $J=0.9\pi$ in the FSPT phase (blue dots) where it exhibits saturating period-two oscillations, $J=0.5$ in the thermal phase (purple diamonds) where $\mathcal{Z}(t)$ immediately averages to zero, and for $J=0.1\pi$ in a trivial MBL phase (orange, triangles) where $\mathcal{Z}(t)$ saturates to a non-oscillating value. Each point is averaged over all initial states and $100$ disorder realizations.
    \label{suppfig:fluxecho_ed} 
        }
\end{figure}

Based on general properties of MBL and SPT systems, we argue that the long-time behavior of $\mathcal{Z}(t)$ can sharply distinguish trivial MBL, FSPT, symmetry-breaking, and thermal behaviors. In particular, we will argue below that:
\begin{align}
	\lim_{t\rightarrow\infty}\lim_{L\rightarrow\infty}
	\overline{\mathcal{Z}(t)} = 
	\begin{cases}
		1 & \text{ trivial-MBL} \\
		(-1)^t & \text{ $\Z_2$-FSPT} \\
		0 & \text{ symmetry-breaking MBL} \\
		0 & \text{ thermal}
	\end{cases}
	\label{eq:loschmidt}
\end{align}
where $\overline{(\dots)}$ denotes disorder averaging. Finite size corrections to these formulas are $\mathcal{O}(e^{-L/\xi})$ where $\xi$ is the localization length. We also note that thermal and symmetry-breaking MBL can be distinguished by the latter having non-vanishing $\overline{~|\mathcal{Z}(t)|~}$. This predicted behavior is confirmed by numerically exact diagonalization (ED) simulations of the FSPT model (see Fig.~\ref{suppfig:fluxecho_ed}).

\paragraph{Intuitive picture} Intuitively, we can understand the FSPT result as follows: the $\Z_2$-FSPT is characterized by a quantized pumping of symmetry parity (henceforth called ``charge")~\cite{potter2016classification,roy2016abelian,kumar2018string}. Namely, during each period, for a system with closed, periodic boundary conditions an odd number of symmetry charges encircle the system. In the presence of a symmetry flux, the $\Z_2$ analog of the Aharonov-Bohm effect dictates that a unit charge encircling a $\pi$-flux acquires phase $(-1)$. Otherwise, the local LIOM dynamics is insensitive to the global flux sector (to see this note that for any spatially-well localized operator, one can always choose a gauge such that the local action of flux insertion is equivalent to a physically-inconsequential gauge transformation). Hence, in $t=n$ periods, the evolution with- and without- the flux will then differ by $(-1)^n$.

\paragraph{Locally-implementable approximation} We then construct a strictly-local approximate flux-insertion operator, $F$, which has the property that, in MBL phases, $F|\psi\>$ has finite overlap with the exact flux-inserted state: $|\psi_F\>$. Denoting by $\tilde{\mathcal{Z}}(t)$ the local-approximation obtained by evaluating $\mathcal{Z}(t)$ with the exact flux insertion operator replaced by its local approximation: $\U_F \rightarrow F^\dagger \U F$, we argue that $ \overline{~\tilde{\mathcal{Z}}(t)~} \sim c \cdot\overline{\mathcal{Z}(t)} $ where $0<c<1$ is model-dependent constant of order $\log c \sim -\xi$, i.e.~that the local approximation to the Loschmidt flux echo tracks the true value with finite fidelity. We further design and implement a circuit to measure $\tilde{\mathcal{Z}}$ for the FSPT model utilizing an ancillary qubit. Of course, the experimental implementation suffers from the same limitations due to coherent errors described in the main text, and yields good agreement with simulations only up to $\sim 5$ Floquet periods.

\subsection{Formalism}
\subsubsection{Gauging the symmetry}
Following a standard ``gauging" procedure, we define a formal symmetry-gauging procedure in two steps. First, one expands the Hilbert space to include gauge-link variables $\v{\tau}_{i,i+1}$ on each nearest neighbor bond where $\tau^z$ represents a $\Z_2$-gauge connection (the discrete analog of the Wilson line segment $e^{i \int \v{A}\cdot d\v{r}}$ of electromagnetism), and $\tau^x$ represents the $\Z_2$ gauge-electric field. Second, one projects into the subspace of this enlarged Hilbert space that obeys the Gauss' law constraint: $G_i = 1~\forall i$ with:
\begin{align}
G_i = \tau^x_{i-1,i}\sigma^x_i\tau^x_{i,i+1},
\end{align}
(the discrete and lattice analog of Gauss' law $\nabla\cdot \v{E}(\v{r})=\rho(\v{r})$ for electromagnetism). Crucially, we consider $\tau$-variables to be completely non-dynamical ``background" gauge variables, i.e.~which have Hamiltonian $=0$.

Any $\Z_2$-symmetric local Hamiltonian can be similarly gauged by adding a connected string of $\tau^z$'s connecting between every pair of symmetry-charged  single site operators ($\sigma^{y,z}$) in each term of the Hamiltonian~\footnote{Note that there are guaranteed to be an even number of these factors in any term of a symmetric Hamiltonian. In $1d$, imposing strictly locality in all terms removes any ambiguity for different choices of pairings.}.

For example, the $\Z_2$ FSPT model in the main text can be generated by a local time-dependent Hamiltonian which changes as follows under the gauging procedure:
\begin{align}
H &= \sum_i \[J_i(t)\(X_iX_{i+1}+Y_iY_{i+1}\) + h_i(t)X_i\] 
\nonumber\\
& \hspace{1.2in}\text{\red{gauge}} \Big\downarrow 
\nonumber\\
H_G=&\sum_i \[J_i(t)\(X_iX_{i+1}+Y_i\red{\tau^z_{i,i+1}}Y_{i+1}\) + h_i(t)X_i\],
\end{align}
where $J_i(t),h_i(t)$  the (piecewise constant) time-dependent coefficients that reproduce the unitary circuit dynamics shown in Fig.~\ref{fig:fspt}a.

\begin{figure*}[t!] 
    \centering
    \includegraphics[width=1.0\textwidth]{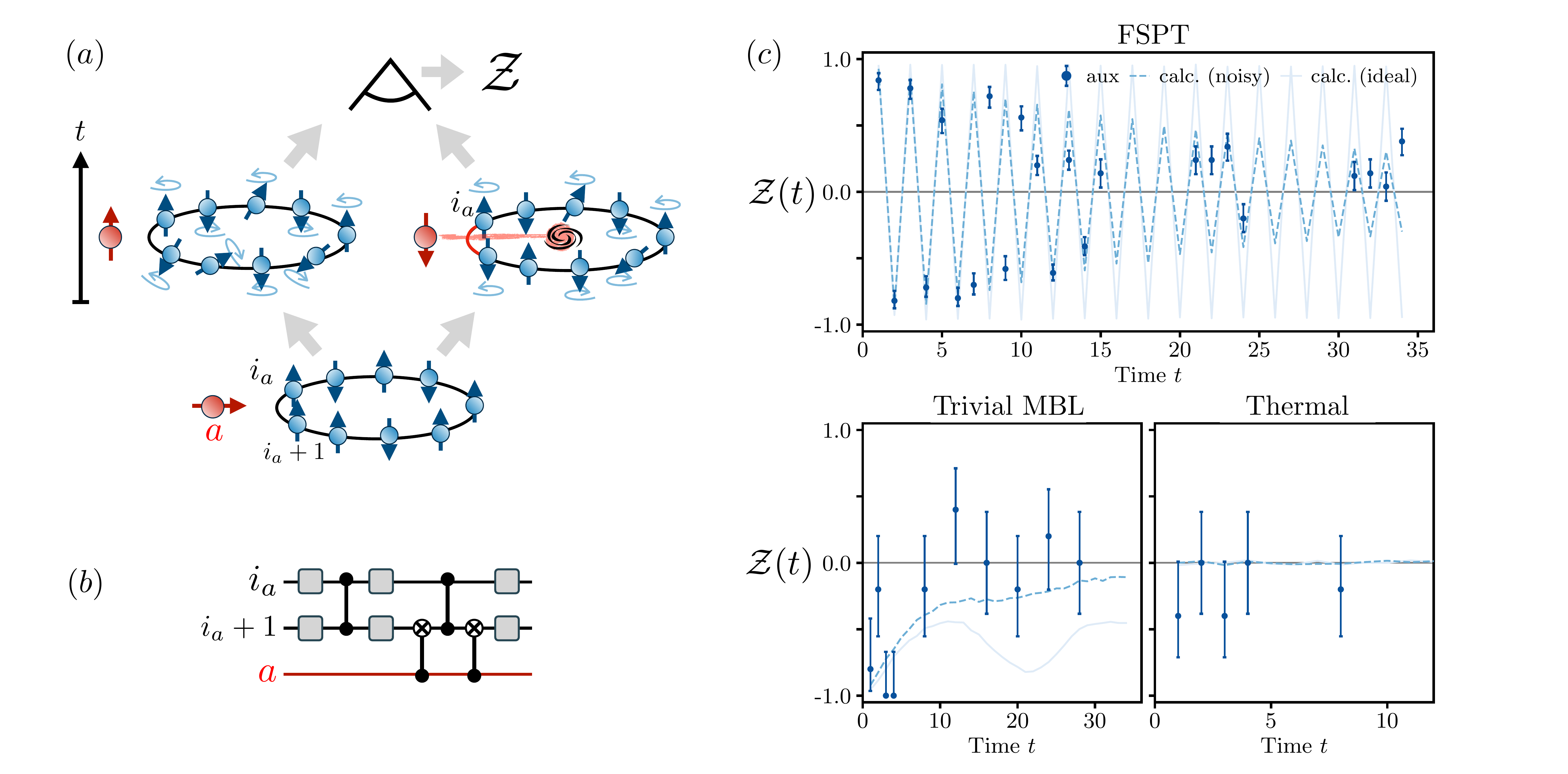}
    \caption{ {\bf Loschmidt flux echo}
    \label{suppfig:fluxecho} 
    (a) Interferometric protocol for measuring the Loschmidt flux-echo, $\mathcal{Z}(t)$: controlled evolution and measurement of an ancilla , $a$, shown here in red enables measurement of the overlap between time evolved states with and without a symmetry flux inserted. (b) Circuit for implementing the ancilla-controlled flux insertion. (c) Simulated and experimental data for the Loschmidt flux echo in the Floquet model with $J=0.9\pi$ (top panel) in the FSPT phase, $J=0.1\pi$ (bottom left panel) in the trivial MBL phase, and $J=0.5\pi$ (bottom right panel) in the thermalizing phase. For ideal noiseless simulations (pale line, simulation), symmetry-preserving MBL $\mathcal{Z}(t)$ exhibits persistent period-two oscillations in the FSPT phase, saturates to a constant value in the trivial MBL phase, and rapidly decays in the thermalizing phase. For weakly-open but symmetry-preserving MBL systems (dashed line, simulation) the oscillation or saturation amplitude slowly decays due to incoherent errors. By contrast, the experimental data strongly deviates from the symmetric simulations after $t\gtrsim 5$ Floquet periods due to coherent errors. Experimental data shown for trivial-MBL and thermal include only ten shots, and were taken for a different disorder realization than the FSPT data.
        }
\end{figure*}
\subsubsection{Formal Flux Insertion}
After projection, the system retains $2^L$ (gauge-invariant) degrees of freedom consisting of: $2^{L-1}$ spin configurations that satisfy the charge-neutrality constraint implied by Gauss' law: $\prod_i \sigma^x_i = \prod_i \tau^{x}_{i-1,i}\tau^{x}_{i,i+1}=1$, and a global $\Z_2$ ``magnetic" flux: $\Phi = \prod_i \tau^z_{i,i+1} = \pm 1$. This foreshadows that we will be able to effectively simulate the gauged system without introducing extra qubits to represent the $\tau$ variables.

We can formally define the action of a flux-insertion operator, $F$, on any local gauge-invariant operator $\mathcal{O}_i$ with bounded support, centered at site $i$ as follows: Choose a reference bond, $(i_a, i_a+1)$ at which to insert the flux, and divide the system (with periodic boundary conditions) into an interval $A$ of length $L/2$ centered at $i_a$ and its complement, $B$ centered at the antipodal point $i_b=(L/2-i_a)\text{ mod }L$. Then define the ``flux-inserted" version of $\mathcal{O}$ as:
\begin{align}
\mathcal{O}(x)\rightarrow \mathcal{O}_F(x) &= 
\begin{cases}
	f_+^\dagger \mathcal{O}_{F,i} f_+ & x\in A \\
	\mathcal{O}(x) & x\in B
\end{cases}, 
\nonumber\\
f_+ &= \(\prod_{x_b<x<x_a}\sigma^x_{i}\) \tau^x_{i_a,i_a+1}.
\end{align}
For any operator in $A$ away from site $i_a$, $f_+$ acts like a pure gauge transformation. Furthermore, this definition is actually independent of the reference points $i_{a,b}$ (up to a gauge transformation). While the above definition holds sharply for strictly local operators (with finitely-bounded support), one can extend it to exponentially-well localized operators with the minor caveat that the definitions depend on choice of $i_{a,b}$ with sensitivity $\sim \mathcal{O}(e^{-L/2\xi})$.

As an application, we can use this definition to identify pairs of symmetry-preserving gauged-MBL eigenstates in different flux sectors, and show that each element of the pair has nearly identical energy to accuracy $O(e^{-L/\xi})$. Following standard practice, we define a Floquet system as being MBL if its Floquet operator (time evolution for a single time-step/circuit layer) can be written as:
\begin{align} 
	\U = W^\dagger \Xi e^{-iE[\sigma^x_i]} W 
\end{align}
where $E[\sigma^x_i]$ is an exponentially-well localized function of $\sigma^x_i$'s, and $W$ is a (symmetric) finite-depth local unitary that implements the weak-local dressing of single-site spins $\sigma^x_i$ into conserved local integrals of motion (LIOMs):
\begin{align}
	\ell_i = W^\dagger \sigma^x_i W.
\end{align}
In the gauged system (with non-dynamical background gauge fields), the gauged analog of $W$ (defined by applying the gauging procedure to the local generator of $W$: $-i\log W$, and which, in a slight notational abuse, we will denote by the same symbol as the ungauged version) depends on the gauge-variables only through $\tau^z$ , i.e.~$W$ commutes with the gauge-flux operators. Here, $\Xi$ is a symmetric operator satisfying $[\Xi,E]=0$. For all the phases we consider here, $\Xi^2=1$. For example, in the gauged $\Z_2$ FSPT with periodic boundary conditions:
 \begin{align}
 	\Xi_\text{FSPT} = \Phi = \prod_i \tau^z_{i,i+1},
\end{align}
is the $\Z_2$ gauge flux operator (with open boundary conditions, $\Xi$ would capture the quantized spin echo dynamics of the topological edge states). 

The key property of symmetric MBL is that the (quasi)-energy eigenstates can be uniquely specified by listing the eigenvalues of the LIOM operators. In a gauged system, for each state in the even flux sector: $|\{s_i\},\Phi=+1\>$ defined by $\ell_i|\{s_i\},\Phi=+1\>=s_i|\{s_i\},\Phi=+1\>$,  there is a partner state in the odd gauge-flux sector: $|\{s_i,\Phi=-1\> = F|\{s_i\},\Phi=+1\>$ defined by $\ell_{F,i}|\{s_i,\Phi=-1\> = s_i|\{s_i,\Phi=-1\>$ (i.e.~labeled by the same LIOM eigenvalues but with flux-inserted LIOM operators $\ell_{F,i}$). Crucially, since the action of $F$ is locally equivalent to a gauge transformation everywhere, the partner states have the same (quasi)-energy, $E_F$, (up to $\mathcal{O}(e^{-L/\xi})$ corrections).

From these definitions and observations, one can readily verify Eq.~\ref{eq:loschmidt}. In particular, for the FSPT:
\begin{align}
	\mathcal{Z}(t=n) &= \frac{\text{tr}~\U^\dagger_F(n) \U(n)}{\text{tr} \mathbbm 1} \\
	&= \frac{\text{tr}~W^\dagger F^\dagger \Phi^n F e^{+inE_F} W  W^\dagger \Phi e^{-inE} W}{\text{tr} \mathbbm 1}
	\nonumber\\
	&= \frac{1} {\text{tr} \mathbbm 1}~
	\underset{(-1)^n}{\underbrace{\(F^\dagger \Phi^n F \Phi^n\)}}
	 \underset{1+\mathcal{O}\(e^{-L/\xi}\)}{\underbrace{\(W^\dagger e^{+i(E_F-E)}W\)}} 
	\nonumber\\
	&= (-1)^n+\mathcal{O}\(e^{-L/\xi}\)
\end{align}
where we have used that $[W,\Phi]=0$ since $W$ is diagonal in the $\tau^z$ basis. 

\paragraph{Behavior of flux-echo in other phases} Similar arguments show that a trivial MBL phase ($\Xi=\mathbbm{1}$) would have $\mathcal{Z}(t)_\text{trivial-MBL}=1$. By contrast, in a thermal phase the localization length diverges, and the system becomes sensitive to the global flux (e.g. the ratio of the Thouless energy to level spacing diverges~\cite{serbyn2015criterion}), and to exhibit chaotic response, such that the evolutions with and without flux insertions deviate exponentially with time, resulting in exponential decay of $\mathcal{Z}$. 

Finally, we note that in MBL systems with spontaneous symmetry breaking (SSB), the above arguments fail. For example, there is not quite a complete set of symmetry-preserving LIOMs. For example, if we consider a ferromagnetic spin glass, a natural set of LIOMs would be a locally dressed version of $\{\sigma^z_{i}\sigma^z_{i+1}\}$, which count the parity of domain walls in the magnetic order, but require one additional integral of motion for completeness (e.g. either a local asymmetric operator like $\sigma^z_i$ or a symmetric, but non-local operator such as $\prod_i\sigma^x_i$). In the gauged system, the total parity of domain walls is locked to the gauge flux in the system, so that there is a local energy cost to inserting flux, and the quasi-energies with and without flux differ by an $O(1)$ constant. As a result, for SSB-MBL, $\mathcal{Z}(t)\sim e^{-i\e t}$ where $\e$ depends on the location of flux insertion and disorder configuration. In particular, the disorder average $\overline{\mathcal{Z}(t)}=0$ at long times, but the average modulus remains $\overline{|\mathcal{Z}(t)|}=1$. The latter property distinguishes the SSB-MBL behavior from that of thermal systems for which $\overline{|\mathcal{Z}(t)|}=0$.

Together, these behaviors show that the flux-echo provides a complete non-local ``order parameter" diagnosing all possible thermal and MBL phases with symmetric drives. This property makes the flux-echo a potentially useful diagnostic. For example, a previously introduced non-local string-order parameter of the FSPT phase~\cite{kumar2018string} could not distinguish between FSPT and SSB-MBL requiring measurement of multiple order parameters to uniquely diagnose the phase.

\paragraph{Local approximation to flux-echo} While the above arguments demonstrate the existence of a formally exact flux-insertion procedure, in practice, this procedure requires explicit knowledge of LIOMs. However, we now argue that it is sufficient to locally approximate the action of flux insertion without knowledge of the LIOMs, which enables a practical measurement scheme for the flux-echo. 

Since the LIOMs, $\ell_i$ of a symmetric MBL system are related by local dressing, $W$ to single-site spin operators $\sigma^x_i$, one can write the exact flux insertion operator $F = W^\dagger\tau^x_{i_a} W$ (where, recall that $i_a$ is an arbitrarily chosen location of flux-insertion). While the exact flux insertion operator $F$ cannot be implemented without knowledge of $W$, one can approximate the action of $F$ by simply acting with $\tau^x_{i_a}$ which differs only by quasi-local, and symmetric dressing from $F$, and its correlation function:
\begin{align}
	\tilde{\mathcal{Z}}(t) = \frac{\text{tr}~\tau^x_{i_a} \U^\dagger(t) \tau^x_{i_a} \U(t)}{\text{tr}~\mathbbm{1}}
\end{align}
would hence have finite overlap with the exact flux Loschmidt echo, $\mathcal{Z}(t)$. Approximating the action of $W$ as randomly scrambling operators within a localization length $\xi$, generically, we expect the disorder averaged magnitude of this quantity $\overline{|\tilde{\mathcal{Z}}(t)|}$ to be smaller than the $\overline{|\mathcal{Z}(t)|}$ by a finite constant factor of order $0<c \sim e^{-\xi} <1$.

These arguments are supported by numerical simulations of the FSPT model (see Fig.~\ref{suppfig:fluxecho_ed}).

\subsection{Quantum Circuit Implementation}
The flux-echo witness $\tilde{\mathcal{Z}}$ can be measured interferometrically (see Fig.~\ref{suppfig:fluxecho}) using an ancilla qubit initialized in an equal superposition of $\frac{1}{\sqrt{2}}\(|0\>+|1\>\)$, to control the time-evolution, such that the system evolves under $H(t)$ or $\tau^x_{i_a}H(t)\tau^x_{i_a}$ when the ancilla is in the $|0\>$ or $|1\>$ state respectively. For the FSPT model, this amounts to simply allowing the ancilla qubit to flip the sign of a single $e^{-iJ\theta \sigma^y_{i}\sigma^{y}_{i+1}}\rightarrow e^{+iJ\theta \sigma^y_{i}\sigma^{y}_{i+1}}$, which can be accomplished with only two extra CNOT gates per Floquet period as shown in Fig.~\ref{suppfig:fluxecho}b. Then, measuring $\<\sigma^{x,y}\>$ for the ancilla reveals the real or imaginary parts of $\tilde{\mathcal{Z}}$ respectively. 

We have implemented this protocol both in classical simulations and experimentally, with results shown in Fig.~\ref{suppfig:fluxecho}. The experimentally measured flux-echo in the FSPT phase initially shows oscillations that survive up to $\approx 5$ Floquet periods, but then are dephased due to coherent errors (as indicated by the oscillations with large amplitude, but wrong phase compared to numerical simulations). We note that this dephasing occurs $\approx 2\text{-}3\times$ more rapidly than that for the OBC circuits, suggesting that the more complicated ion transport paths required to implement periodic (ring) boundary conditions exacerbate the effects of errors, consistent with the expected results of spatial inhomogeneities in the magnetic-field as discussed above. In contrast, in the thermal phase the flux-echo immediately dies, and in the trivial MBL phase the data is consistent with a saturation to a constant value up to an overall slow decay due to weakly-open MBL effects.

\section{Long-time (meta)stability of recursively-generated quasiperiodic drives} \label{AppendixMagnus}
This section addresses the long-time fate of the EDSPT model, in an idealized perfectly isolated system. We will see that the EDSPT behavior is not a rigidly stable phase, but rather a very-long-lived but ultimately metastable behavior. However, the lifetime for the edge modes can be made exponentially long in a certain parameter regime, and hence can easily be made much longer than any experimental lifetime.

Topologically-trivial paramagnetic and time-quasicrystalline behavior arising in such recursive Fibonacci drives were previously studied theoretically and numerically in~\cite{dumitrescu2018logarithmically}, leveraging the exponential growth of $t_n$ with number of recursions $n$ to efficiently reach exponential long times in simulations\footnote{Asympotically $t_n \sim \varphi^n$ where $\varphi=\frac{1+\sqrt{5}}{2}$ is the golden ratio.}. There, it was observed that, for strong-disorder, MBL only occurs as a meta-stable phenomena in these recursively driven systems: rather than saturating to finite constant, spin-autocorrelators decayed logarithmically slowly, with MBL dynamics eventually melting away into infinite temperature incoherent at ultra-long `heating time-scales $t_h\sim e^{1/\delta}$ where $\delta$ represents the drive strength (more precisely, deviation from an idealized perfectly-commuting drive). Numerical simulations (Fig.~\ref{suppfig:heating}) indicate that this long-lived MBL behavior also arises in the EDSPT model. 

Additionally, in this appendix we extend a recursive adaptation of high-frequency (Magnus-type) expansion to show that the dynamics are governed by an effective time-independent Hamiltonian with emergent dynamical symmetries, for times up to $t_*\sim \delta^{-p}$ with $p=3,5$ for topologically non-trivial (see below) and trivial~\cite{dumitrescu2018logarithmically} phases respectively. The long-time dynamics between $t_*$ and $t_h$ appears to be beyond the purview of \emph{any} effective Hamiltonian description, and a controlled theoretical description of this regime remains elusive. However, the recursive nature of the drive permits efficient numerical access to long-time dynamics, and our simulations indicate that the topological edge spins and emergent dynamical symmetries persist well beyond $t_*$ up to $t_h$.

\subsection{Recursive Magnus expansion}
In Ref.~\cite{dumitrescu2018logarithmically}, we previously developed a recursive high-frequency (``Magnus") expansion technique to reduce the evolution under a weak quasiperiodic drive generated by a Fibonacci sequence of two circuit layers to an effective time-independent Hamiltonian evolution. This technique was accurate when the unitary of the generating circuit layers was close to the identity by an amount $\sim \delta$. The resulting effective Hamiltonian accurately captured the evolution up to time $t_*\sim \delta^{-5}$. Comparison to numerical simulations showed that, beyond $t_*$, the driven system exhibited logarithmically slow heating causing complete thermalization in timescale $t_h\sim e^{1/\delta}$.

Here, we generalize this technique to the case relevant for the idealized EDSPT model, i.e.~when generating unitaries $\U_{x,z}$ are near-perfect $\pi$-pulses about the $x,z$ axes respectively. I.e.~$\U_{x,z}^2\approx \mathbbm{1}+\mathcal{O}(\delta)$ but $\U_{x,z}$. We will show that, in this parameter regime, one can again obtain an approximate time-independent Hamiltonian description up to times $t_*\sim \delta^{-3}$. Furthermore, the resulting effective Hamiltonian has an emergent pair of $\Z_2$ (Ising) symmetries whose generates $g^{\alpha}$ with $\alpha \in\{x,z\}$ are related to $\prod_{i}\sigma^{\alpha}_{2i}\sigma^{\alpha}_{2i+1}$ by a finite-depth local unitary transformation whose precise form depends on the details of the drive. Our theoretical picture of the EDSPT phase, is that these emergent dynamical symmetries protect the SPT edge modes of an effective AKLT-chain, ``encrypted" in a quasiperiodically rotating frame of the drive.

While this technique provides our best known analytic handle on recursively generated quasiperiodic drives, numerical simulations and experimental results suggest that the recursive Magnus expansion dramatically underestimates the stability and survival time scales for the emergent dynamical symmetries. First, we note that the model implemented in the main text, which has maximal disorder strength $4\pi$ for the K-couplings is actually not in the small $\delta$-regime, yet it still exhibits long-lived signatures of topological edge states. Moreover, based on the  numerical simulations that access long times on modest system sizes observe that these oscillations decay exponentially-slowly with respect to $1/\delta$, suggesting that the emergent dynamical symmetries survive up to this much longer $t_h$ timescale. 

As a preview, the main result of this section is that, up to third order in a small parameter, $\delta$ that quantifies the deviation from some exactly-solvable ideal drive, we can approximately reduce the unitary evolution to the form:
\begin{align}
	\U_{6n} = V^\dagger {\rm e}^{- i \varphi^{6n} \left( D + \dots \right) }  V
\end{align}
where, $D$ is an effective time-independent Hamiltonian that obeys an emergent $\Z_2\times \Z_2$ symmetry generated by $g^{x,z}=V^\dagger \prod_i\sigma^{x,z}_i V$ respectively. Here, $V$ is a finite depth local unitary that we explicitly construct, and  $(\dots)$ includes i) terms with subleading (and generally oscillatory) n-dependence (e.g. with coefficients decreasing as $\sim \varphi^{-6n}$ or faster), and ii) $\mathcal{O}(\delta^3)$ terms beyond the validity of the expansion. We note that, for reasons that will become apparently shortly, it is convenient to express the evolution at Fibonacci-indices that are multiples of $6$, and note that similar expressions can be obtained for $\U_{6n+k}$ with $k \in \{1,\dots 5\}$.

As with the closely-related expansion for topologically trivial Fibonacci drives previously derived in~\cite{dumitrescu2018logarithmically}, this expansion has the peculiar property that it breaks down at finite order independent of expansion parameter $\delta$. Namely, at $O(\delta^3)$, the expansion results in terms in $D$ that grow faster that $\varphi^{6n}$ regardless of $\delta$ and finite system size, whereas on general grounds for sufficiently small $\delta$ one can must always be able to reduce $\U_{6n} = e^{-i\varphi^{6n}D'_n}$ for some bounded, Hermitian $D'_n$ (though not necessarily one that is time- i.e.~n-independent) since the circuit contains only $\sim \varphi^{6n}$ layers. Hence, the generation of terms with coefficients growing faster than $\varphi^{6n}$ signal a breakdown in the recursive Magnus expansion, and hint at a possible obstruction to describing the dynamics beyond $t_*$ by \emph{any} effective time-independent Hamiltonian.

\subsection{Inflation rule}
To derive the recursive Magnus expansion, we exploit a self-similar fractal structure of Fibonacci sequences under ``inflating" the generating unitaries $\U_{x,z}$.
Let us consider a Fibonacci drive generated by two pulses $(\U_x=X {\rm e}^A, \U_z=Z {\rm e}^B)$ where $X=\prod_i \sigma^x_{2i}\sigma^z_{2i+1}$, $Z=\prod_i \sigma^z_{2i} \sigma^z_{2i+1}$ commute and square to one, and $A,B$ are anti-Hermitian operators with small norm $\delta \sim \Vert A,B \Vert \ll 1$ that represent perturbations to the ideal drive. For convenience we also define $Y=XZ$ (note that in our notation $X,Y,Z$ are \emph{not} single spin Pauli operators, but rather strings of even numbers of $\sigma^{x,y,z}$ products).

\begin{widetext}
The Fibonacci drive is generated by the inflation rule:
\begin{align}
\(Xe^{A}, Ze^{B}\) \rightarrow \(Ze^{B}, Xe^{A}Ze^{B}\) = \(Ze^{B},Ye^{A_x}e^B\).
\end{align}
where we have defined $A_x = XAX$, and similarly define $A_{y,z},B_{x,y,z}$. It is convenient to recurse three times with the inflation rule, and pull all $X,Z$'s to the left (conjugating A,B's along the way) until the $X,Z$ prefactors return to their original form:
\begin{align}
\(Xe^{A}, Ze^{B}\) &\overset{(1)}{\rightarrow} \(Ze^{B}, Xe^{A}Ze^{B}\) = \(Ze^{B}, XZe^{A^z}e^{B}\) 
\nonumber\\&\overset{(2)}{\rightarrow}
	\(XZe^{A_z}e^{B}, Ze^{B} XZe^{A_z}e^{B}\) = \(XZe^{A_z}e^{B}, Xe^{B_y} e^{A_z}e^{B}\) 
\nonumber\\&\overset{(3)}{\rightarrow} 
	\(Xe^{B_y} e^{A_z}e^{B}, ~ Ze^{A_y}e^{B_x}e^{B_y} e^{A_z}e^{B}\) .
\end{align}
This ``3x-inflated'' inflation rule yields an recursion relation for A,B alone:
\begin{equation}
({\rm e}^A, {\rm e}^B) \overset{(3x)}{\longrightarrow} ({\rm e}^{B_y} {\rm e}^{A_z} {\rm e}^{B},{\rm e}^{A_y} {\rm e}^{B_x} {\rm e}^{B_y} {\rm e}^{A_z} {\rm e}^{B}),
\end{equation}
 where the arrow now represents this ``3x'' recursion. 
 
 At this point, it turns out to be easier to think of the unitaries: $\U_n$ as being functions of an ``alphabet'' of 8 ``letters": $(A,B,A_x,B_x,A_y,B_y,A_z,B_z )$. To avoid a proliferation of subscripts, in what follows we change our notation to label Fibonacci times with indices that are multiples of $3$, by defining:
 \begin{align}
 	U_{n} \equiv\U_{3n},
\end{align}
where $U_n$ corresponds to $t_{3n} = F_{3n}\sim \varphi^{3n}$ circuit layers.
 The recursion rules for $A_x,A_y,A_z$ and $B_x,B_y,B_z$ follow from those of $A,B$ simply by conjugating by $X,Y,Z$. We have the following recursion relation for the unitary operator:
\begin{equation}
U_{n+1}({\rm e}^A,{\rm e}^B,{\rm e}^{A_x},{\rm e}^{B_x},{\rm e}^{A_y},{\rm e}^{B_y},{\rm e}^{A_z},{\rm e}^{B_z} ) = U_{n} ({\rm e}^{B_y} {\rm e}^{A_z} {\rm e}^{B},{\rm e}^{A_y} {\rm e}^{B_x} {\rm e}^{B_y} {\rm e}^{A_z} {\rm e}^{B}, {\rm e}^{B_z} {\rm e}^{A_y} {\rm e}^{B_x},\dots),
\end{equation}
 where we emphasize that $n$ now labels the number of ``3x'' recursions, and $U_1={\rm e}^A$.

\subsection{Effective Hamiltonian and generalized Magnus expansion} 
Our goal is to check the emergence of an effective Hamiltonian for the dynamics by computing order by order $H_n = \log U_n$ (defined here to be anti-Hermitian), which obeys 
\begin{equation}
H_{n+1}(A,B,A_x,B_x,A_y,B_y,A_z,B_z ) = H_{n} (B_y \star A_z \star B,A_y \star B_x \star B_y \star A_z \star B, B_z \star A_y \star B_x, \dots), \label{eqrecursionH}
\end{equation}
 where $A \star B \star C \star \dots = \log ( {\rm e}^A  {\rm e}^B {\rm e}^C \dots)$. 
 \end{widetext}
 
 To proceed, we relabel those letters $(K_1,K_2, \dots, K_8) = (A,B,A_x,B_x,A_y,B_y,A_z,B_z ) $, and expand  
 \begin{align}
 H_{n} = \sum_{\alpha=1}^8 c_\alpha(n) K_\alpha + \sum_{\alpha \beta} C_{\alpha \beta}(n) \left[K_\alpha, K_\beta \right] +\dots~.
 \end{align}
 where $(\dots)$ indicate nested commutators with larger number of $K's$ that contribute at $\mathcal{O}(\delta^3)$.
 
Plugging this expression in the recursion relation~\eqref{eqrecursionH}, we find that the leading order coefficients are given by:
\begin{equation}
\vec{c}_\alpha(n+1) = \hat{M}^T \vec{c}_\alpha(n),
 \end{equation}
with the recursion matrix 
\begin{equation}
\hat{M} = 
\left( \begin{array}{cccccccc}
0 & 1 & 0 & 0 & 0 & 1 & 1& 0 \\
0 & 1 & 0 & 1 & 1 & 1& 1 & 0 \\ 
0 & 0 & 0 & 1 & 1 & 0 & 0 & 1\\
0 &1 & 0 & 1 & 1 & 0 & 1 & 1\\
0 & 1 & 1 & 0 & 0 & 1 & 0 & 0\\
1 & 1& 1 &0 & 0 & 1 & 0 & 1\\
1 & 0 & 0 & 1 & 0 & 0 & 0 & 1\\
1 & 0 & 1& 1 & 0 & 1 & 0&  1\\
  \end{array} \right).
 \end{equation}
This recursion relation can be solved straightforwardly by diagonalizing $\hat{M}$. The largest eigenvalue is $\varphi^3$, with an eigenvector $(1/\varphi, 1, 1/\varphi, 1, 1/\varphi, 1, 1/\varphi, 1)$. This implies an {\em emergent} symmetry between $A,A_x,A_y,A_z$ (and same for $B,B_x,B_y,B_z$)   at long times. Namely, from this dominant eigenvector coefficients we can see that $c_1 \sim c_3 \sim c_5 \sim c_7$ at large $n$, and $c_2 \sim c_4 \sim c_6 \sim c_8$. In other words, at this order $H_n$ commutes with the pules $X$ and $Z$ at large $n$ (long times). 

Let us now consider the next order terms $C_{\alpha, \beta}$. There are two different types of contributions: one coming from applying the substitution rule $ \tilde{K}_\alpha \to \sum_{\beta} M_{\alpha \beta} K_\beta +\dots$ to leading order to the term $\sum_{\alpha \beta} C_{\alpha, \beta}(n) \left[\tilde{K}_\alpha, \tilde{K}_\beta \right] $ on the right hand side, and the other one coming from the Baker, Campbell, Hausdorff (BCH) formula from $\sum_{\alpha} c_\alpha(n) \tilde{K}_\alpha $ . The first type is readily taken into account, and we find:
\begin{equation}\label{eqRecursionrelationC}
\hat{C}(n+1) = \hat{M}^T \hat{C} (n)\hat{M}+ \hat{r}_n, 
 \end{equation}
where $\hat{r}_n$ is a skew-symmetric matrix that originates from the second type of terms mentioned above. 

Keeping track of the contributions to $\hat{r}_n$ is straightforward albeit cumbersome. For example, we have a term $c_1(n)  \tilde{K}_1 = c_1 \tilde{A}$ with $\tilde {A} \to B_y \star A_z \star B $. This will generate terms $\frac{c_1(n)}{2} \left[ B_y, A_z \right] + \frac{c_1(n)}{2} [B_y,B] + \frac{c_1(n)}{2} [A_z,B]$ that should be included in $ \hat{r}_n$. The contributions from $c_3(n), c_5(n), c_7(n)$ can be obtained by conjugating by $X,Y,Z$, and the other contributions can be dealt with in a similar way (except there are now five exponentials to expand using the BCH formula). With this expression for $\hat{r}_n$, eq.~\eqref{eqRecursionrelationC} can be solved by going to the eigenbasis of $\hat{r}_n$ and then rotating back. The explicit expression for the matrix $\hat{C} (n)$ is not particularly illuminating, but we have checked that it grows with $n$ as $\varphi^{3n}$, corresponding to time, which means that one can write $H_n = -i \varphi^{3n} (D + \dots)$, where $D$ can be interpreted as an effective Hamiltonian for the dynamics. We will show that $H_n$ has an {\em emergent} ${\mathbb Z}_2 \times  {\mathbb Z}_2$ symmetry at large $n$ (long times), which protects the topological edge modes.   

This expansion breaks down at the next order: including nest commutators in the expression of $H_n$, we find that the coefficients of such terms grow with $n$ exponentially faster than $\varphi^{3n}$. Not only does the Hamiltonian interpretation therefore break down at this order, but higher orders become more important earlier in time: this appears to be a general property of recursive drives as noted in Ref.~\cite{dumitrescu2018logarithmically}. This means that this high-frequency, Magnus-type expansion is only strictly valid for times up to $t_\star \sim \delta^{-3}$, with $\delta \sim \Vert A,B \Vert \ll 1$. 

However, as previously commented, we find numerically that even away from this high-frequency regime, strong disorder and MBL protect this dynamical phase  up to much longer, exponential time scales (see Fig.~\ref{suppfig:heating}). 

\begin{widetext}
\subsection{Emergent symmetry}

Let us now analyze the symmetries of $H_n$ including second order terms. Let 
\begin{align}
\hat{g}_x &= 
\left( \begin{array}{cccccccc}
0 & 0 & 1 & 0 & 0 & 0 & 0& 0 \\
0 & 0 & 0 & 1 & 0 & 0& 0 & 0 \\ 
1 & 0 & 0 & 0 & 0 & 0& 0 & 0 \\ 
0 & 1 & 0 & 0 & 0 & 0& 0 & 0 \\ 
0 & 0 & 0 & 0 & 0 & 0& 1 & 0 \\ 
0 & 0 & 0 & 0 & 0 & 0& 0 & 1 \\ 
0 & 0 & 0 & 0 & 1 & 0& 0 & 0 \\ 
0 & 0 & 0 & 0 & 0 & 1& 0 & 0 \\ 
  \end{array} \right), 
~~~~ \hat{g}_z = 
\left( \begin{array}{cccccccc}
0 & 0 & 0 & 0 & 0 & 0 & 1& 0 \\
0 & 0 & 0 & 0 & 0 & 0 & 0& 1 \\
0 & 0 & 0 & 0 & 1 & 0 & 0& 0 \\
0 & 0 & 0 & 0 & 0 & 1 & 0& 0 \\
0 & 0 & 1 & 0 & 0 & 0 & 0& 0 \\
0 & 0 & 0 & 1 & 0 & 0 & 0& 0 \\
1 & 0 & 0 & 0 & 0 & 0 & 0& 0 \\
0 & 1 & 0 & 0 & 0 & 0 & 0& 0 \\
  \end{array} \right).  
 \end{align}
This is a representation of ${\mathbb Z}_2 \times {\mathbb Z}_2$ on the 8-dimensional ``alphabet", whose action corresponds to conjugating letters by $X$ and $Z$, and $\hat{g}_x^2 = \hat{g}_z^2=1$ and $[\hat{g}_z,\hat{g}_z]=0$. Note that the recursion matrix is compatible with this symmetry:
\begin{equation}
[\hat{M}, \hat{g}_z] = [\hat{M}, \hat{g}_x] = 0.  
\end{equation}
The leading order coefficients $c_\alpha(n)$ are symmetric only at long times, as the initial condition $c_\alpha(n=0) = \delta_{\alpha,1}$ breaks the symmetry. Because of this, the matrix $\hat{r}_n$ is also ``almost'' symmetric up to ${\cal O}(1)$ errors: $[\hat{r}_n, \hat{g}_{x,z}] =  {\cal O}(1)$. However, these non-symmetric terms get amplified in the recursion relation~\eqref{eqRecursionrelationC}, so  $\hat{C} (n)$ contains some exponentially large non-symmetric terms of order ${\cal O}(\varphi^{3n})$. We will write 
\begin{equation} \label{eqNonSymm}
\hat{C} (n) = \hat{C}_s (n) + \delta {\hat C}(n),
\end{equation}
with the symmetrized matrix $\hat{C}_s = (\hat{C}+ \hat{g}_x\hat{C} \hat{g}_x + \hat{g}_z\hat{C} \hat{g}_z + \hat{g}_x \hat{g}_z\hat{C} \hat{g}_x \hat{g}_z)/4$, and $||\delta {\hat C}(n)|| \sim{\cal O}(\varphi^{3n})$.

Let us now try to cancel out those symmetric term by a (finite-depth unitary) change of frame:
\begin{equation}
V = {\rm e}^{\frac{1}{2}\sum_\gamma v_\gamma K_\gamma},
\end{equation}
with $U_n' = \hat{V} U_n \hat{V}^\dagger$.

We find
\begin{equation}
H'_{2n} = \log U'_{2n}= \frac{1}{2}\sum_{\gamma \alpha} c_\alpha(2n) v_\gamma \left[ K_\gamma, K_{\alpha}\right] +  \sum_{\alpha} c_\alpha(2n) K_\alpha + \sum_{\alpha \beta} C_{\alpha \beta}(2n) \left[K_\alpha, K_\beta \right] +\dots,
\end{equation}
 where we restricted ourselves to even times to avoid odd/even effects. Our goal is to choose $v_\gamma$ to cancel out the non-symmetric terms $\delta {\hat C}(n),$ in~\eqref{eqNonSymm}, or more precisely, cancel out the leading non-symmetric terms $\delta {\hat C}(2n) = \varphi^{6n} \left( \delta {\hat C}_0 + \dots\right)$,  where the dots represent exponentially decaying terms. We find that this can be achieved by choosing $v_\alpha = \left(-\frac{2 + \sqrt{5}}{4} -  \frac{1 - \sqrt{5}}{2} c, c,
-\frac{\sqrt{5}}{4} -  \frac{1 - \sqrt{5}}{2} c, c, \frac{1}{4} - \frac{1 - \sqrt{5}}{2} c, c,  \frac{3}{4} -  \frac{1 - \sqrt{5}}{2} c,  c-1 \right)$ with $c$ a constant. Note that this is only possible since the leading order terms $\sum_{\alpha=1}^8 c_\alpha(n) K_\alpha$ in $H_n$ are symmetric at large $n$. 
\end{widetext}

We conclude that for this choice of dressing unitary, we can write
\begin{equation}
U_{2n} = V^\dagger {\rm e}^{- i \varphi^{6n} \left( D + \dots \right) }  V,
\end{equation}
 with an effective Hamiltonian $ D$ that has a ${\mathbb Z}_2 \times {\mathbb Z}_2$ symmetry generated by the pulses, and where the dots denote exponentially small in $n$ non-symmetric terms. In other words, the unitary evolution operator $U_{2n}$ commutes with the ``dressed'' symmetry generators
 \begin{equation}
g_x = V^\dagger  X V, \ g_z = V^\dagger  Z V,
 \end{equation}
 at long times. This symmetry is manifestly {\em emergent}, as $V$ depends on $A$ and $B$. The effective Hamiltonian has the following expression
\begin{align}
- i D =& \frac{1}{\sqrt{5}} \left(B_s + \varphi^{-1} A_s \right) +  \frac{3 \sqrt{5} - 5}{40} \left[A,B \right]_s +\nonumber\\
&+  \frac{\sqrt{5} - 5}{40} \left[A,B_x \right]_s +\frac{1}{4\sqrt{5}}\left[A,B_z \right]_s.
\end{align}
 with $A_s = (A+A_x +A_y +A_z)/4$, and similarly for all symmetrized quantities, {\it e.g.} $\left[A,B \right]_s = ( \left[A,B \right] +  \left[A_x,B_x \right] +  \left[A_y,B_y \right]+  \left[A_z,B_z \right])/4$.

\subsection{Numerical (ED) Simulations}
\begin{figure}[tbh] 
    \centering
    \includegraphics{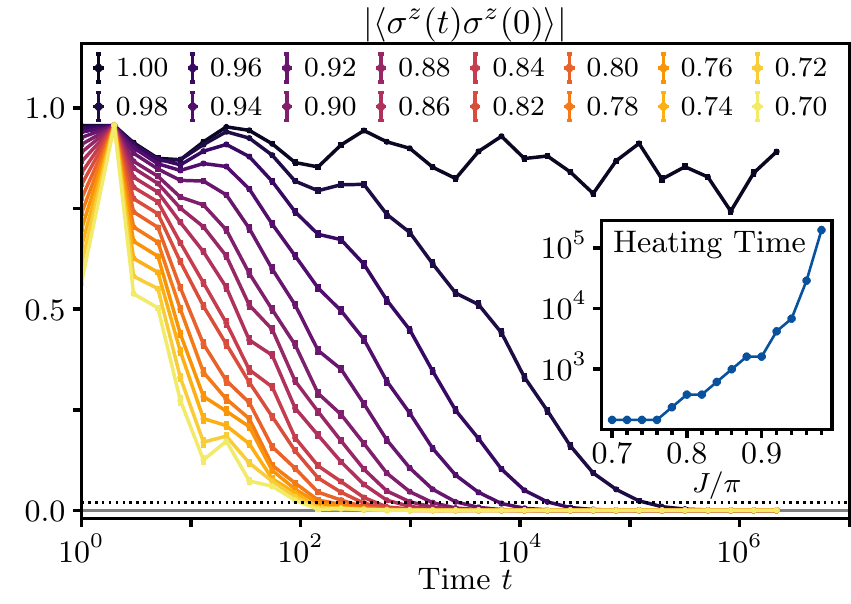}
    \caption{{\bf Slow heating (numerical simulations)}  Numerical simulations of the long-time behavior of $|C_z|$ for the edge spin of an $L=10$ spin-chain with the Fibonacci drive, for various $J$ (other parameters chosen as in Fig.~\ref{fig:qp}). Each point is averaged over $960$ disorder realizations. The inset shows estimate of heating time $t_h$ empirically defined as the time where $|C_z|$ drops below a small threshold $\eps = 0.02$ (dashed line), and demonstrates that the three-fold periodic oscillations persist to time scale exponentially long in the inverse pulse detuning, $|J-\pi|^{-1}$.
    \label{suppfig:heating} 
        }
\end{figure}
To investigate the long-time dynamics beyond the regime of validity of the recursive Magnus expansion, we have performed numerically exact simulations of the EDSPT model for various $J$. The recursive structure enables access to exponentially long times $t_n\sim F_n$ with $n$ recursions. Fig.~\ref{suppfig:heating} shows the resulting edge-spin correlators, averaged over disorder, for various $J$ values. For $J$ close to $\pi$, we observe that the quasiperiodic oscillations of the edge spins survive up to very long times. We estimate the heating time $t_h$ as the time at which the edge-oscillation drops below an arbitrary small threshold: $\eps=0.02$. As shown in the inset of Fig.~\ref{suppfig:heating}, $t_h$ shows superpolynomial growth $1/\delta = \pi/|J-\pi|$ for small $\delta$, indicating that the edge spin oscillations survive \emph{far beyond the time-scale at which recursive Magnus expansion fails}. This behavior is similar to that numerically observed in topologically-trivial recursive Fibonacci drives~\cite{dumitrescu2018logarithmically}, and suggests that ultra-long-lived but ultimately metastable MBL-like dynamics are generic features for recursive drives with strong disorder. However, a detailed theoretical picture of the logarithmically slow entanglement growth and heating in these models remains elusive at this time.

\section{Additional data} \label{app:additionaldata}
In this appendix, we include additional simulation and experimental data for the FSPT and EDSPT models.

\subsection{Short time FSPT dynamics}
\begin{figure}[t] 
    \centering
    \includegraphics[width=\columnwidth]{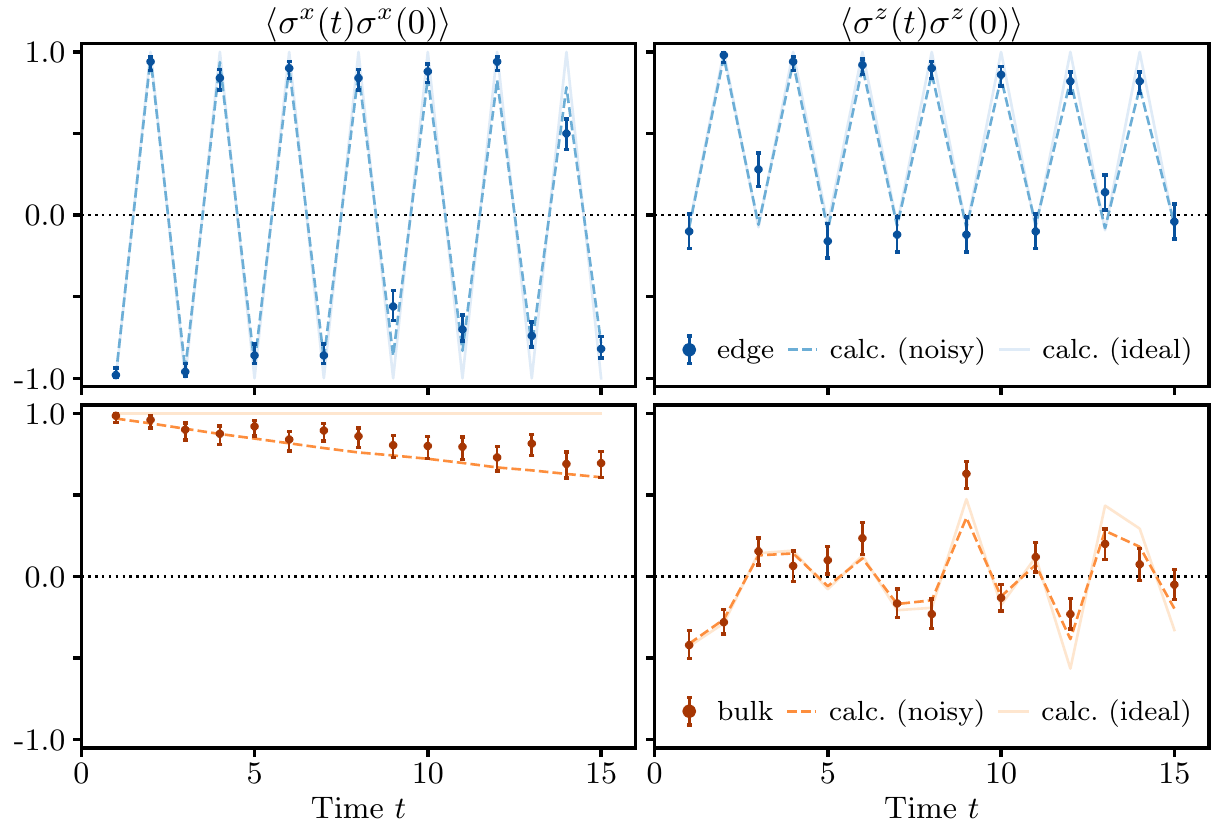}
    \caption{{\bf Ideal FSPT implementation (short times) } FSPT model simulation (noiseless: pale solid lines, noisy: dashed lines) and experimental data (solid points with error bars) at the ideal $J=\pi$ ``fixed-point" for short times $t\leq 15$ for which coherent errors have not yet affected the topological edge spins. The edge shows the expected period-two oscillations, whereas bulk spins show slowly decaying, weakly-open MBL behavior along the symmetry axis $\sim\sigma^x$, and rapidly dephase due to random fields and spin-spin interactions perpendicular to the symmetry axis, $\sim\sigma^z$.
    \label{suppfig:fsptshort}
        }
\end{figure}

Figure~\ref{suppfig:fsptshort} shows a close-up of the early time dynamics of the FSPT model for nominally ideal $J=\pi$ parameters, which, absent errors, would be at the fixed point of the FSPT phase. During the time interval shown the system is not yet effected by coherent errors and exhibits the expected weakly-open MBL FSPT dynamics, characterized by period two edge oscillations and random/dephasing bulk oscillations.

\subsubsection{EDSPT with uniaxial disorder and nominal $\Z_2$ symmetry}
\begin{figure*}[h!] 
    \centering
    \includegraphics{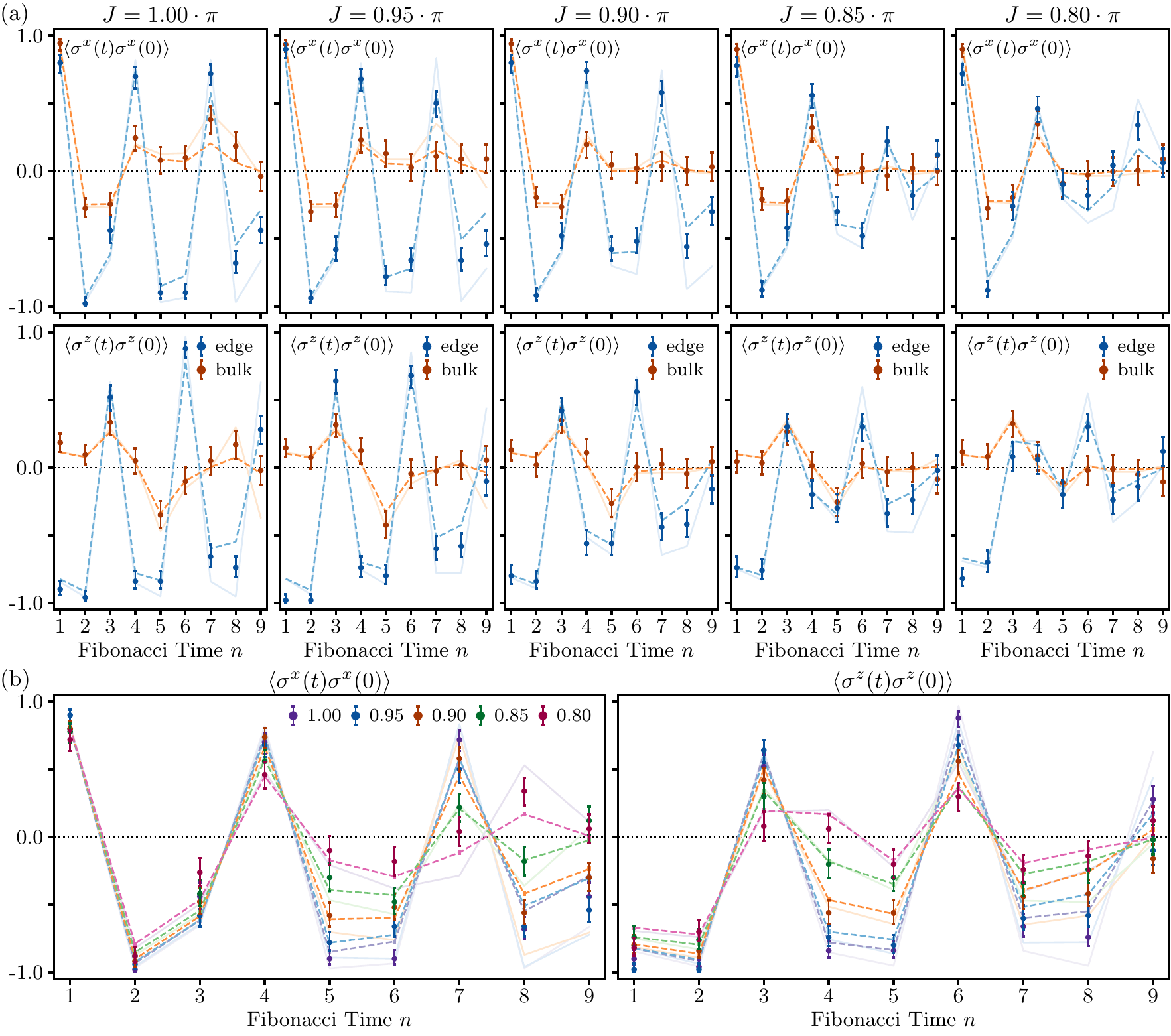}
    \caption{{\bf EDSPT with uniaxial disorder. } (a) Edge and (site averaged) bulk correlators for various values of $J$  for the EDSPT model with $\Z_2$ symmetry ($\v{B}_i\parallel \hat{y}$), which show that the EDSPT behavior persists over a range of $|J-\pi|/\pi \lesssim 0.25$. Pale lines show idealized (noiseless) simulations, dashed lines show simulations with depolarizing noise, and solid dots with $1\sigma$-error bars are experimental data. (b) The same edge data for various $J$ replotted on a single plot to facilitate comparison of curves with different pulse-weights.
    \label{suppfig:z2dis_1}
        }
\end{figure*}

\begin{figure*}[h!] 
    \centering
    \includegraphics{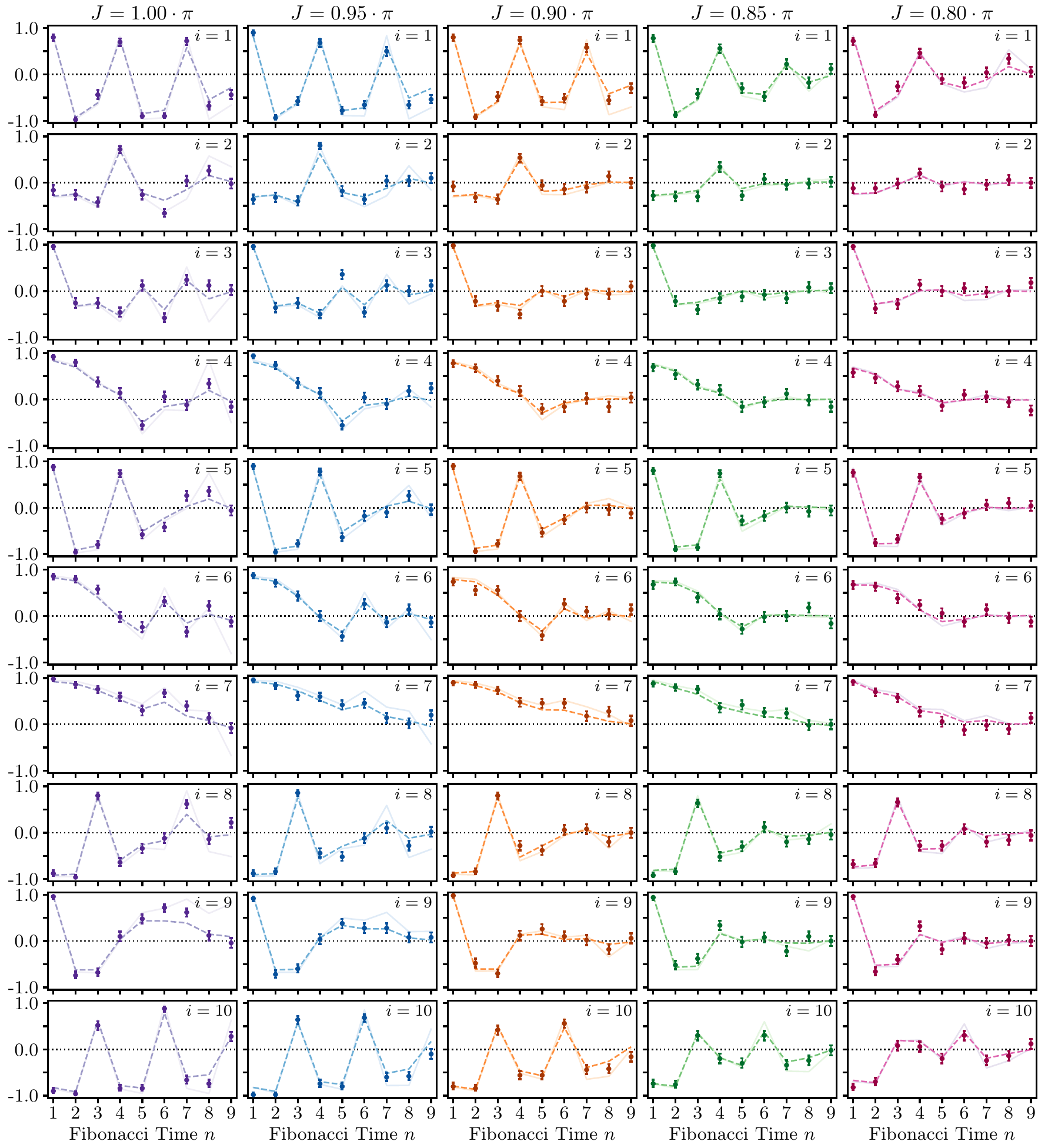}
    \caption{{\bf EDSPT Site-resolved Correlators (with uniaxial disorder)} shown for each site $i\in \{1,2,\dots 10\}$ in the chain. Pale lines show idealized (noiseless) simulations, dashed lines show simulations with depolarizing noise, and solid dots with $1\sigma$-error bars are experimental data. As described in the Methods section, even sites are prepared and measured in the $z$ basis and odd sites in the $x$ basis. Whereas edge sites ($i=1,10$) exhibit period-three oscillations in Fibonacci time, the bulk sites ($2\leq i\leq 9$) undergo random oscillations that wash out upon site- or disorder- configuration averaging.
    \label{suppfig:z2dis_2}
            }
\end{figure*}

This section shows results for the EDSPT model with random fields purely along the y-axis, which, like the FSPT model, nominally has a microscopic $\Z_2$ symmetry generated by $\prod_i \sigma_i^y$, but which is broken in implementation by the same coherent errors that decohere the FSPT edge states. The plots shown in Fig.~\ref{suppfig:z2dis_1} below provide further evidence that the EDSPT edge states are not harmed by these coherent errors and do not rely on this fine-tuned symmetry, a fact that is accentuated by the data shown in the main text in which this symmetry is manifestly broken by the vector $\v{B}$-disorder.  Additionally, we show the raw, non-averaged bulk-data in Fig.~\ref{suppfig:z2dis_2}. Note that the bulk spins exhibit random, but slowly decaying oscillations characteristic of the slow dephasing dynamics of MBL. Upon disorder- and/or site- averaging these random oscillations wash out, as seen in the other figures presented throughout the main text. 

\clearpage

\end{document}